\documentclass[11pt,a4paper]{article}

\usepackage{amssymb}
\usepackage{amsmath}
\usepackage{float}
\usepackage{url}
\usepackage{amsthm}
\usepackage[sort]{cite}

\theoremstyle{remark}

\usepackage{dsfont}
\usepackage{xcolor}
\usepackage{graphicx}
\usepackage{bm}
\usepackage[margin=1in]{geometry}
\usepackage[colorlinks=true,citecolor=magenta,linkcolor=blue]{hyperref}

\newcommand{\be}{\begin{equation}}
\newcommand{\ee}{\end{equation}}
\newenvironment{equations}{\equation\aligned}{\endaligned\endequation}
\newcommand{\fer}[1]{(\ref{#1})}

\newcommand{\eps}{\varepsilon}

\newcommand{\R}{\mathbb{R}}
\newcommand{\N}{\mathbb{N}}

\newcommand{\fc}{\widehat f}

\newcommand{\Bf}{\mathbf{f}}
\newcommand{\bQ}{\mathbf{Q}}
\newcommand{\bR}{\mathbf{R}}

\title{From Individual-Based Models to Macroscopic Dynamics of Antimicrobial Resistance} 
\date{}

\author{
		Marco Menale\thanks{\texttt{marco.menale@unina.it}}\\
		{\small	Department of Mathematics and Applications} \\
		{\small University of Naples, Italy} \\
		\\
		Giuseppe Toscani\thanks{\texttt{giuseppe.toscani@unipv.it}}\\
		{\small	Department of Mathematics ``F. Casorati'', University of Pavia, Italy} \\
		{\small  Institute of Applied Mathematics and Information Technology ,  Pavia, Italy} \\
		\\
		 Mattia Zanella\thanks{\texttt{mattia.zanella@unipv.it}} \\
		{\small	Department of Mathematics ``F. Casorati''} \\
		{\small University of Pavia, Italy}
		}

\begin{document}
\maketitle
\begin{abstract}
We introduce and discuss a kinetic framework describing the time evolution of the statistical distributions of a population divided into the compartments of susceptible, infectious, recovered, and resistant in the presence of a microbial infection driven by susceptible–infectious interactions. Our main objective is to quantify the impact of excessive and inappropriate antimicrobial use, which accelerates the spread of resistance by enabling a fraction of infectious individuals to transition into the resistant compartment.
The model consists of a system of Boltzmann-type equations capturing binary interactions between susceptible and infectious individuals, complemented by linear redistribution operators that represent recovery, the development of resistance, and reinfection processes. In the grazing-collision limit, we show that this Boltzmann system is well approximated by a system of coupled Fokker–Planck equations. This limiting description allows for a more tractable analysis of the dynamics, including the characterization of the long-time behavior of the population densities.
Our analysis highlights how interaction terms drive the system toward a stable equilibrium and quantifies the effects of inappropriate antimicrobial use on the distribution of resistant individuals. Overall, the results offer a multi-scale perspective that bridges kinetic theory with classical epidemic modeling.\\


\end{abstract}

\section{Introduction}

Antimicrobial resistance (AMR) has emerged as a persistent global public health challenge, with projections estimating up to 10 million deaths annually by 2050. AMR occurs when viruses, bacteria, fungi, or parasites no longer respond to antimicrobial treatments in humans or animals, enabling these microorganisms to survive and proliferate within the host. The primary driver of this crisis is the excessive and inappropriate consumption of antimicrobials—particularly antibiotics—which continues to accelerate the global spread of resistance.

The scale of the antimicrobial resistance crisis has been highlighted by several reports from the World Health Organization (WHO), which identify AMR as one of the most serious global public health threats. According to recent estimates, antimicrobial resistance was directly responsible for approximately 1.27 million deaths worldwide in 2019 and contributed to nearly 4.95 million deaths, placing it among the leading causes of mortality globally. The burden of AMR affects countries across all income levels, although its impact is particularly severe in low- and middle-income regions where healthcare infrastructures and surveillance systems are often 
limited.\footnote{World Health Organization (2023): {Antimicrobial Resistance} \eject \url{https://www.who.int/news-room/fact-sheets/detail/antimicrobial-resistance}}

Beyond its direct health consequences, antimicrobial resistance threatens many advances of modern medicine. Procedures such as major surgery, cancer chemotherapy, and organ transplantation become significantly riskier in the presence of resistant pathogens. Moreover, AMR entails substantial economic costs: projections suggest that by 2050 the global burden associated with resistant infections could reach trillions of dollars, accompanied by significant reductions in global gross domestic product. The consequences of antimicrobial resistance extend well beyond mortality. Resistant infections typically require longer and more complex treatments, resulting in prolonged hospital stays, increased healthcare expenditures, and additional pressure on already strained healthcare systems. Furthermore, the spread of resistance disproportionately affects vulnerable populations, exacerbating existing inequalities in access to effective medical care. Taken together, these factors highlight that antimicrobial resistance is not only a biomedical challenge but also a major socio-economic issue, with profound implications for public health systems and global development.

These considerations underscore the urgent need to better understand the mechanisms governing the emergence and spread of resistance, as well as to develop effective strategies to mitigate its impact. For these reasons, antimicrobial resistance has often been described as a {silent pandemic}, a term used by international organizations, e.g. United Nations,\footnote{{United Nations Regional Information Centre} (2024): {The Global Threat of Antimicrobial Resistance: A Silent Pandemic}, \eject \url{https://unric.org/en/the-global-threat-of-antimicrobial-resistance-a-silent-pandemic/}} to underline the progressive and often underestimated spread of resistant infections. Unlike acute epidemic outbreaks, the impact of AMR accumulates gradually over time through increasing treatment failures, longer hospitalizations, and higher mortality rates. This slow but persistent evolution makes antimicrobial resistance particularly difficult to control and highlights the need for coordinated global strategies.

In a worst-case scenario \cite{o2016tackling}, antimicrobial resistance could become one of the leading causes of death worldwide, potentially accounting for up to $10$ million deaths annually. Such projections would rank AMR among the foremost causes of mortality, surpassing major contributors such as cancer and road traffic accidents. In addition to its dramatic impact on human health, the economic consequences could be equally severe: global cumulative costs have been estimated to reach up to $100$ trillion, alongside significant reductions in worldwide economic productivity and gross domestic product.

Mathematical modeling plays a crucial role in understanding the mechanisms behind the emergence and transmission of AMR, and it provides a foundation for exploring and designing innovative control strategies. Ensuring that such models are broadly applicable requires adherence to sound modeling practices \cite{birkegaard2018send,HANDEL2009655}. Today, mathematical models are widely recognized as essential decision-support tools in medicine and public health \cite{opatowski2011contribution}, significantly advancing our understanding of how AMR develops, emerges, and spreads.

A common approach in this field is the use of compartmental models, which aim to capture key mechanisms of resistance transmission or to explore hypothetical scenarios \cite{spicknall2013modeling,weinstein2001understanding,caudill2016role,lopez2000modelling,eftimie}. Most of these models follow the classical framework introduced by Kermack and McKendrick \cite{kermack1927}, who proposed a seminal description of epidemic spread by partitioning the population into susceptible (S), infectious (I), and recovered (R) compartments.
Variants of these SIR-type models can incorporate additional features, such as reinfection dynamics, through appropriate extensions \cite{hethcote2000}.
 
These models provide valuable conceptual insight into disease transmission based on empirical evidence or heuristic assumptions. Indeed, since these models are not derived from first principles, they inherently simplify reality, often neglecting heterogeneity in individual interactions.
To address these limitations, mathematical research has increasingly focused on refining the description of temporal changes within compartments, especially in systems characterized by multiple time scales. Methods from kinetic theory has offered a promising framework for describing the evolution of individuals across compartments through statistical distributions over additional variables that capture population features, such as in-host heterogeneity, contact rates, or evolving behavioral traits \cite{dimarco2021kinetic,zanella2023,bertaglia2023,della2023sir,lorenzi24,BisiLorenzani,bertaglia24,pulvirenti_simonella,Pulvi,bellomopredicting}. This approach also enables the study of spatially dependent epidemic dynamics \cite{albi2022kinetic,ha22,Difrancesco25}.

Despite these advances, relatively few results exist on rigorously connecting microscopic trajectories to macroscopic averaged quantities. Significant research efforts across various communities have focused on deriving mean-field models, which describe the collective behavior of large populations of interacting agents and can be obtained through rigorous limiting procedures \cite{bolley12,carrillo10,degond08,ha09,motsch14}. While microscopic models suffer from the curse of dimensionality, mean-field formulations retain essential emergent features while enabling the study of macroscopic patterns and phase transitions \cite{during09,during15,MR4251319,MR3067586,preziosi}.

In a recent work \cite{MTZ},  a mean-field formulation of a general compartmental epidemic model of the form
\be\label{vec-LV}
\frac{\partial\Bf(x,t)}{\partial t} =  \bQ(\Bf(x,t)) + \bR(\Bf(x,t)); \qquad  \Bf(x,t=0) = \Bf_0(x),
\ee	
has been introduced and studied. 
In \fer{vec-LV} the non-negative variable $x \in \mathcal I \subseteq \mathbb R_+$ measures the size of the populations in the various compartments.  Here $\bQ(\Bf(x,t))$ is the $n$-dimensional vector whose components $Q_K( \Bf)$, $K \in \mathcal C$ are suitable (generally nonlinear) kinetic  operators which describe the variation of the $n$-dimensional vector $\Bf(x,t)$ of the densities $f_K(x,t)$ of individuals  in the $K$-th compartment due  to interactions with individuals
in other compartments, while  $\bR(\Bf(x,t))$ is the $n$-dimensional vector whose components $R_K( f_K)$, $K \in \mathcal C$ are linear kinetic redistribution operators which quantify the variation of the  densities $f_K(x,t)$, $K \in \mathcal C$ due to passage towards and/or arrival from other compartments \cite{bondesan2025lotka,TosZan,toscani_zanella26}. System \fer{vec-LV} was built to be closely related to the classical epidemiological model by Kermack and McKendrick \cite{kermack1927}, which in this picture represents  the time evolution of the mean values $m_K(t)$ of the densities $f_K(x,t)$,  $K \in \mathcal C$, i.e. 
\[
m_K(t) = \int_{\R_+} xf_K(x,t) \, dx.
\]
This approach offer a quantitative framework for assessing how population heterogeneity influences the eventual steady state of an epidemic. They also provide a principled basis for calibrating external interventions, which act by modifying the initial population densities and thereby shaping the long-term behaviour of the system.

In this paper, extending the approach of \cite{MTZ} we incorporate the effects of antimicrobial resistance consistently with the AMR dynamics explained in\cite{bonhoeffer1997evaluating,bergstrom2004ecological,lehtinen2017evolution}. Mathematical models of AMR are numerous and typically enrich classical epidemic frameworks by introducing additional compartments to capture resistance-related dynamics \cite{opatowski2011contribution,spicknall2013modeling,mulberry2020systematic,HANDEL2009655,GRUNNILL2022111277}.
More precisely, we consider the  infectious compartment is divided into two subpopulations: the compartment $I$, representing individuals infected with a sensitive strain, and the compartment $J$, representing individuals infected with a resistant strain. This refinement reflects the assumption that the pathogen can occur in two forms, depending on its resistance to treatment.
In addition, we partition the group of treated individuals into two compartments: the compartment $T$, corresponding to successfully treated individuals originating from $I$ when treatment is effective, and the compartment $R$, representing individuals undergoing treatment associated with resistant infections. The mean-field formulation \fer{vec-LV} of the model accounts for the fact that individuals in the compartment $R$ may originate from both infectious classes, namely from $J$, as well as from $I$ in cases of inappropriate or ineffective treatment. This mechanism constitutes the key feature of the present work, as it allows us to investigate the impact of improper antimicrobial use on the emergence and spread of resistance.

Furthermore, we analyse the emergence of equilibrium states and investigate the impact of transitions to the resistant compartment induced by inappropriate treatments. Within the proposed framework, the misuse of antimicrobial treatments appears to exert a stronger influence on the emergence and spread of resistance than the beneficial effects associated with well-targeted therapeutic strategies.

The paper is organized as follows. Section \ref{sec:kinetic} is devoted to a precise description of the elementary interactions which enable to build up this kinetic model of antimicrobial resistance, which is dealt with in Sections \ref{subsec:comp} and \ref{sec:inte}.  The main feature of this model, which describes the evolution of the population's distributions in terms of their territorial densities by means of a system of bilinear Boltzmann--type equations, briefly derived in  Appendix \ref{sec:appendix}, is its connection with more classical descriptions of antimicrobial resistance in terms of a system of ordinary differential equations, in the spirit of the classical SIR model of Kermack and McKendrick \cite{kermack1927}, a description which appears when evaluating the mean values of the statistical distribution of the involved populations. Then, Section \ref{sec:FP} introduces a simplified version  of the Boltzmann  type kinetic model, in the form of a system of Fokker--Planck type equations with variable in time coefficients of diffusion and drift. The large--time evolution of the solution to the system of Fokker–Planck type equations is subsequently studied in Section \ref{sec:large}. Last, numerical simulations illustrating the effectiveness of the approach are presented in Section \ref{sec:numerics}.


\section{A kinetic description of antimicrobial resistance}\label{sec:kinetic}

\subsection{The compartments}\label{subsec:comp}
As briefly discussed in the Introduction, we consider a population split into five groups of individuals, from now on defined  compartments, characterized by their state of health $K \in \mathcal C$. First, we have susceptible individuals, say $S$, that are disease-free. Then, the classical group of infectious is here split into two further compartments, $I$ and $J$. Specifically, the compartment $I$ is composed by infectious individuals characterized by a sensitive-strain pathogen, while the compartment $J$ is characterized by infectious individuals with a resistant-strain pathogen. This refined description hypothesizes that the involved pathogen can be present in two different forms, depending on its resistance to treatments: a sensitive-strain and a resistant strain. We further assume that the infectious individuals can be treated through different medical actions that may lead to different consequences. Consequently,  the group of treated individuals is divided in two compartments: the compartment $T$ of the well-treated individuals, coming from the compartment of infectious $I$ when the pathogen does not develop  resistance to the right medical treatment, and the compartment $R$   of resistant-treated individuals, where we find individuals that can come either from the group $J$ or from the group $I$, in case individuals were not treated in a right way. This latter assumption is the most relevant feature of this current paper, since it tries to understand the consequences of a wrong use of medical treatments. Thus, we aim to study the evolution in time of the statistical densities of the following five compartments:

\begin{enumerate}
    \item[$i)$] Susceptible $S$;
    \item[$ii)$] Infectious with sensitive strain pathogen $I$;
    \item[$iii)$] Infectious with resistant strain pathogen $J$;
    \item[$iv)$] Well treated $T$ from infectious group $I$;
    \item[$v)$] Resistant treated $R$ coming either from the resistant group $J$ or from the wrong-treaed infectious group with sensitive strain pathogen $I$.
\end{enumerate}

These compartments define the class $\mathcal C =\{S,I,J,T,R \}$. The evolution of the system of densities further requires that individuals in the class $\mathcal C$ can move from one compartment to the other according to precise transition rules, listed below.

\begin{itemize}
    \item Susceptible individuals $S$ can move to either the group $I$ or $J$ through classical infectious processes. The infection can be caused by interactions with individuals belonging to one of the groups $I$, $J$, and $T$. It is worth noticing that the infections related to interactions with the latter group $T$, are very typical in hospital settings.
    
    \item Infectious individuals $I$ with sensitive strain pathogen can be well-treated, and in this case move to the compartment $T$, or, if they are not well-treated,   move to the compartment $R$.
    
    \item Infectious individuals $J$ with resistant strain pathogen can move to the compartment $R$, or, if their infection becomes sensitive strain, they can move into the compartment $I$. 
    
    \item Last, assuming that the reinfection process can occur, individuals in the groups $T$ and $R$ can move to the group of susceptible individuals $S$.
\end{itemize}

\begin{figure}[h!]
    \centering
    \includegraphics[width=120mm]{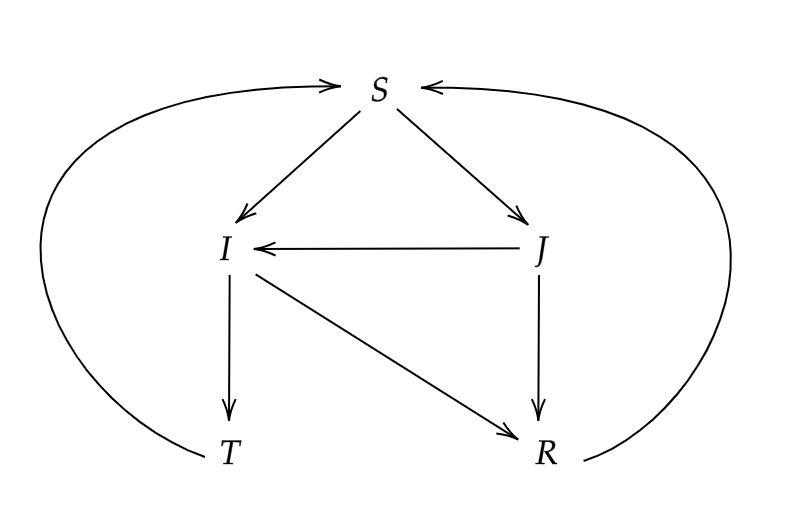}
    \caption{Compartment system}\label{fig1}
\end{figure}

The structure of the possible transitions of individuals from one compartment in the class $\mathcal C$ to another is represented in figure \ref{fig1}.

Let us now briefly introduce the kinetic description of the system, starting from the elementary interactions between individuals. As usual in this context \cite{ParTos-2013}, we assume that individuals in any compartment of $\mathcal C$ are characterized by their statistical distribution with respect to the variable $x \in \mathbb{R}_+$, which quantifies the territorial density of the group once measured with respect to some unit. This choice relies on the fact that the population density is an important factor in epidemic spreading, since it is closely related to the number of contacts of individuals, and consequently to the number of infections. 

We denote by $f_K(x,t)$, $K \in \{S,\, I,\, J,\,T,\,R\}$, the distribution function of the population $K$. Hence, for any given small size interval $dx$,  $f_K(x,t)dx$ represents the fraction of individuals in population $K$ with a density that belongs to the interval $(x,\, x+dx)$ at time $t\geq 0$. We further assume that the statistical distributions $f_K$ are probability densities, so that
\begin{equation}
    \int_{\mathbb{R}_+}f_K(x,t)\,dx=1,\qquad  K \in \mathcal{C}.
\end{equation}
This choice allows to evaluate from the knowledge of the densities $f_K(x,t)$ the principal moments of order $n \in \N$, that for any time $t \geq 0$, are given by
\[
m_K^{(n)}(t) := \int_{\mathbb{R}_+} x^n f_K(x,t)\,dx.
\]
The case $n = 1$ defines the time dependent mean value of the group $K$. In this case, to simplify notations, we define, for $K\in \mathcal{C}$,
\be\label{eq:m}
m_K(t) = m_K^{(1)}(t) := \int_{\mathbb{R}_+} x f_K(x,t)\,dx.
\ee
The case $n = 2$, through the notion of variance,
\begin{equation}\label{eqvar2}
    v_K(t) := \int_{\mathbb{R}_+} \bigl(x - m_K(t)\bigr)^2 f_K(x,t)\,dx
= m_K^{(2)}(t) - [m_K(t)]^2,
\end{equation}
 is useful to quantify the spreading of the size of the populations around their mean values. It is important to note that identity \fer{eqvar2} allows one to express moments of order two in terms of the variance and of moments of order one
\[
m_K^{(2)}(t) = v_K(t) + [m_K(t)]^2.
\]
It is important to remark that, while the evolution of the overall epidemiological system follows the structure described in Section \ref{subsec:comp} and represented in Figure \ref{fig1}, the kinetic description of passages from one compartment to another can be divided into the two  well-separated classes of linear and bilinear elementary interactions. More precisely, linear interactions describe variation of the density in the compartments due to interactions with an external background. On the contrary, bilinear interactions describe variations of the densities in the compartments due to binary interactions between individuals in different classes. 

\subsection{Binary interactions}\label{sec:inte}

 Bilinear interactions are related to the process of infection, and involve the compartments of susceptible individuals $S$ and infectious individuals, belonging to compartments $I$, $J$ or $R$. According to the original description of a predator--prey interaction considered in \cite{bondesan2025lotka}, subsequently adapted to the epidemic spreading in \cite{MTZ}, we assume that individuals in the class $S$ can move to one of the two classes of infectious $I$ and/or $J$ according to the rule
 \be\label{eq:SK}
x^*=x-\phi(y)x+\eta(y)x
 \ee
The elementary rule \fer{eq:SK} implies that the size $x\in \R_+$ of the susceptible individuals changes in time because of his contacts with one of the infected populations $K$ (with $K\in\{I,J,R\}$) of size $y\in\R_+$  according to two different reasons, which quantify the deterministic and random variations. 

As far as the deterministic variation is concerned, it is assumed that it is proportional to the size $y$ of the infected population according to the function
\be\label{eq:phi}
\phi(y) = \beta\frac y{1+y}, \quad 0 <\beta <1,
\ee
where  the constant $\beta$ indicates the maximal rate of infection of the susceptible individuals. Clearly, since in reality this rate of infection can vary in dependence of various unpredictable reasons, it is important also to take into account the contribution of randomness.  To this extent, the random term appearing in \fer{eq:SK} is obtained by assuming that $\eta(y)$  is a random variable of mean value equal to zero, and a variance proportional to the size $y$ of the infectious population
\be\label{eq:SK-var}
 \langle\eta^2(y)\rangle=\sigma\frac y{1+y}, \quad \sigma >0,
\ee
where $\langle X\rangle$ denotes the mean value of the random variable $X$.  The constant $\sigma$ quantifies the maximal rate of spreading of the random variations due to the presence of infectious individuals.
Note that, in absence of infectious individuals $y=0$,  and the size of the susceptible population does not change. On the contrary, to avoid the possibility that the population of susceptible S could disappear ($x^* \le 0$), one has to guarantee that the variable $\eta(y)$ do not take too big negative values. A sufficient condition to prevent this unrealistic behavior is to assume that
\be\label{eq:neg}
- \eta(y) \le 1- \beta.
\ee
Under condition \fer{eq:neg} we have that $x^* >0$, and, as $y>0$ 
\be\label{eq:decreasing}
\langle x^*\rangle = x - \phi(y) x < x,
\ee
which shows that \fer{eq:SK} implies that, in presence of infectious individuals, the size of the population of susceptible  is decreasing in mean. 
However, while essential for the positivity of $x^*$, condition \fer{eq:neg} does not imply smallness conditions on the value of the constant $\sigma$, which in principle can be chosen in such a way that the variance of $\eta(y)$ is very big even in presence of a very small size $y$ of the population of infectious. This weak aspect can be improved by including a condition which guarantees that  $\langle (x^*)^2\rangle \le x^2$.
Since
\[
x^* =(1-\phi(y) -\eta(y))x
\]
the upper bound on the second moment of $x^*$ holds provided
\[
\left\langle (1-\phi(y))^2 +\eta(y)^2 -2\eta(y)(1-\phi(y)) \right\rangle \le 1,
\]
that, since $\langle \eta(y)\rangle =0$, is equivalent to satisfy the inequality
\[
\beta^2\left( \frac y{1+y}\right)^2 +\sigma\frac y{1+y} \le 2\beta\frac y{1+y},
\]
and this inequality holds independently on $y$ provided
\be\label{eq:small-var}
\sigma \le 2\beta -\beta^2.
\ee
Note that, if \fer{eq:small-var} holds, the random effects can be chosen to grow in agreement with the grow of the maximal rate of infection.

The previous discussion can be extended to characterize the growth of infected populations by interaction with the population of susceptibles.
In this case, we assume that the compartment  $K\in\{I,J\}$, of size $y \in\R_+$, can increase  for the entry of individuals from the compartment $S$  according to the rule
 \be\label{eq:K+}
y^* =y+\phi(x)y+\eta(x)y
 \ee
where the deterministic contribution is given by \fer{eq:phi}, the random variable $\eta(y)$  is of mean value equal to zero, and a variance proportional to the size $x$ of the  population $S$ according to \fer{eq:SK-var} and \fer{eq:small-var}.
 
 Of course, interactions with different groups can be characterized by different constants, still subject to bounds \fer{eq:neg} and \fer{eq:small-var}. 
 
 To exemplify, the elementary variation of the group of the compartment $S$ of  susceptible individuals with size $x_S$  due to  interaction with the compartments $I$ of infectious individuals with size $x_I$, and the corresponding variation of the compartment $I$ is described by the interaction pair 
\begin{equations}\label{eq:11}
    x_S'&=x_S-\phi_I(x_I)x_S+\eta_{SI}(x_I)x_S\\
    x_I'&=x_I+\phi_I(x_S)x_I+\eta_{I}(x_S)x_I
\end{equations}
where, for $x \in\R_+$
\be
\phi_I(x)=\beta_I\frac{x}{1+x}\label{eqmicro14}
\ee
and the random variables with zero mean are characterized by
\be\label{eqmicro16}
    \langle\eta^2_{SI}(x)\rangle=\sigma_{SI}\frac{x}{1+x}, \quad
    \langle\eta^2_{I}(x)\rangle=\sigma_{I}\frac{x}{1+x}.
\ee
Same relations for the variation of the pair $(x_S'', x_J'')$ of the pair $(x_S,x_J)$ relative to the $S,J$ interaction. 

A different situation occurs when a susceptible individual with microscopic state $x_S$ could be infected by an individual of the compartment $R$, with microscopic state $x_R$, but this interaction still leads to a passage from $S$ to $J$. In this case we have a passage from the compartment of susceptible to the compartment $J$ of infected individuals with resistant strain caused by interaction with an individual of a third compartment, the compartment $R$. In this case, we have

\begin{equations}
    \label{eqmicro2}
    x'''_S&=x_S-\phi_{R}(x_R)x_S+\eta_{SR}(x_R)x_S\\
    x''_J&=x_J+ \phi_{R}(x_S)x_R+\eta_{RJ}(x_S)x_J.
\end{equations}
In \fer{eqmicro2} we assumed that the random variation of the number of infected in the compartment $J$ is proportional to its number of individuals, with a coefficient proportional to number of individuals belonging to the compartment $R$ according to  the usual law
\begin{align}\label{eqmicro21}
    \phi_R(x)&=\beta_R\frac{x}{1+x}.
\end{align}
The random variables with zero mean are characterized by
\be\label{eqmicro22}
    \langle\eta^2_{SR}(x)\rangle =\sigma_{SR}\frac{x}{1+x}, \quad
    \langle\eta^2_{RJ}(x)\rangle =\sigma_{RJ}\frac{x}{1+x}.
\ee

\subsection{Linear interactions}\label{sec:linear}

Linear interactions describe the passage from one compartment to another due to the presence of the background. The typical situation is the passage from the compartment $I$ of infectious individuals with sensitive strain pathogen to the compartment $T$ of well-treated thanks to correct medical treatment. Following the description in \cite{MTZ}, interactions with a background are modeled as linear redistribution operators \cite{bisi2009}. Let us briefly describe how these linear interactions are built.
Given two compartments $K$ and $L$, with $K,L\in \mathcal C$ and sizes $x_K,x_L \in \R_+$ we consider the linear interaction
\begin{equations}
    \label{eq:redi1}
        x^*_K& = x_K +\gamma(y-\theta x_K),\\ 
      x^*_L &= x_L+\gamma x_K
\end{equations}
which, as clearly stated by the second equality in \fer{eq:redi1}, if $\gamma >0$, moves individuals from $K$ to $L$. As far as the first equality is concerned, the positive constants $\gamma$ and $\theta$ are such that $\theta >1$ and $\gamma\theta <1$, which insures the positivity of the post-interaction state $x^*_K$.
Moreover $y \in \R_+$ is a positive random value distributed according to a probability density $h_{KL}(y,t)$, which represents a background distribution encapsulating for example the availability of medical treatments. This distribution is such that
\be\label{eq:medical}
\int_{\mathbb R_+}yh_{KL}(y,t)dy = (\theta-1)m_K(t),
\ee
where  $m_K(t)$ is the mean value at time $t >0$ of the density  $f_K(x,t)$ of the compartment $K$. Clearly, \fer{eq:medical} guarantees that the  post-interaction value $x'_K$ is decreasing in mean, while 
\be\label{eq:consistaency}
\langle x^*_K + x^*_L\rangle = x_K + x_L.
\ee
We remark that the introduction of a background distribution allows to better clarify from a statistical point of view, the effect of a medical treatment. Tipically, the variance of the background distribution can be easily related to the efficacy of the treatment.

Let us now list below all the situations that can be treated according to this linear interaction. First, infectious individuals with resistant strain pathogen, i.e. with microscopic state $x_J$, can become sensitive strain infectious with microscopic state $x_I$; then, the following microscopic interaction is defined
\begin{equations}\label{eqmicro31}
    x'''_J&=x_J+\gamma_{JI}(y-\theta_{JI}x_J)\\
    x''_I&=x_I+\gamma_{JI}x_J.
\end{equations}
where $y \sim h_{JI}(y,t)$ represent the background such that
\begin{equation}
    \int_{\mathbb{R}_+}yh_{JI}(y,t)\,dy=(\theta_{JI}-1)m_J(t)
\end{equation}
An individual with microscopic state $x_I$, i.e. goes naturally in the compartment $T$ due to the microscopic process
\begin{equations}\label{eqmicro41}
     x'''_I&=x_I+\gamma_{T}(y-\theta_{T}x_I)\\
    x'_T&=x_T+\gamma_{T}x_I,
\end{equations}
where $y \sim h_{IT}(y_2,t)$ represents the background such that

\begin{equation}
    \int_{\mathbb{R}_+}yh_{IT}(y,t)\,dy=(\theta_T-1)m_I(t)
\end{equation}
Otherwise, if the individual is not well treated, it can pass into the compartment $R$, then
\begin{equations}\label{eqmicro51}
    x^{''''}_{I}&=x_I+\gamma_{IR}(y-\theta_{IR}x_I)\\
    x'_{R}&=x_R+\gamma_{IR}x_I,
\end{equations}
where
$y \sim h_{IR}(y,t)$ represents the background such that
\begin{equation}
    \int_{\mathbb{R}_+}yh_{IR}(y,t)\,dy=(\theta_{IR}-1)m_I(t)
\end{equation}
This latter  microscopic process represents the main element in this approach for the antibiotic resistence dynamics.

Finally, the infectious individuals with resistant strain pathogen, i.e. with microscopic state $x_J$, goes into the compartmente $R$, if they do not pass into the compartment $I$. Then, this further microscopic interactions is assumed
\begin{equations}\label{eqmicro61}
    x''''_J&=x_J+\gamma_{R}(y-\theta_{R}x_J)\\
    x''_R&=x_R+\gamma_{R}x_J,
\end{equations}
where
$y \sim h_{JR}(y,t)$ represents the background such that
\begin{equation}
    \int_{\mathbb{R}_+}yh_{JR}(y,t)\,dy=(\theta_{R}-1)m_J(t)
\end{equation}

Last, the loss of immunity of individuals in the compartments $K=T,R$ is described by a linear interaction that takes into account the randomness of the process. Then, we have

\begin{equations}\label{eqmicro71}
    x_L''&=x_L-\alpha_Lx_L+\eta_{LS}x_L\\
    x''''_S&=x_S+\alpha_L x_L,
\end{equations}
where $\eta_{RS}$ and $\eta_{TS}$ are random variables with zero mean and
\begin{equation}\label{eqmicro871}
    \langle\eta_{TS}^2\rangle =\sigma_{TS}(\alpha_T), \quad
    \langle\eta_{RS}^2\rangle=\sigma_{RS}(\alpha_R).
\end{equation}
In particular

\begin{equation*}
    \sigma_{RS}(\alpha_R=0)=\sigma_{TS}(\alpha_T=0)=0,
\end{equation*}
as for $\alpha_R=\alpha_T=0$ the reinfections processes do not occurr. 

\begin{figure}[H]
    \centering
    \includegraphics[width=120 mm]{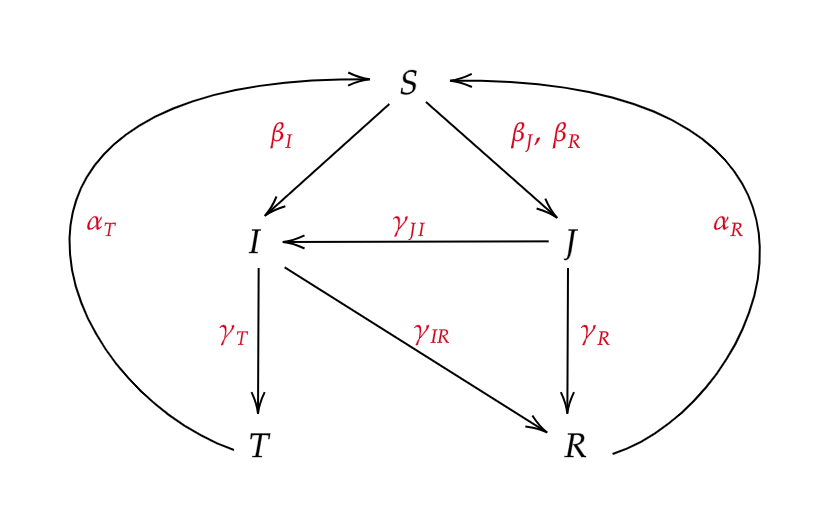}
    \caption{Parameters that quantify the transitions of individuals between compartments }
    \label{fig2}
\end{figure}

Then, bearing Scheme \ref{fig1} in mind, the microscopic processes above described can be read in Scheme \ref{fig2}.

\section{Meanfield description of antimicrobial resistance}\label{sec:FP}

The detailed discussion of Section \ref{sec:kinetic} allows to build a system of kinetic equations of Boltzmann type, where, for $x \in\R_+$, the time variations of the densities $f_K(x,t)$, $K\in \mathcal C$ are determined both by bilinear collision-like operators which take into account the passages from one compartment to another in reason of interactions like \fer{eq:SK} and \fer{eq:K+}, and/or linear operators which describe the passages from one compartment to the other in reason of interactions with a background, as expressed by  \fer{eq:redi1}. 

It is important to note that both binary and linear interactions are built to map positive values into positive values, and that this is obtained by suitably bound the random parts of the interactions themselves. This clearly implies that if the domain of definition of the densities $f_K(x,t)$, $K\in \mathcal C$ at time $t =0$ coincides with $\R_+$, it remains $\R_+$ at any subsequent time $t >0$, so that the mean values $m_K(t),K\in \mathcal C$, defined in \fer{eq:m}  remain strictly positive, unless the densities are concentrated on the value $x=0$. 

The kinetic derivation of the Boltzmann system is  standard, and follows along the lines of Ref. \cite{MTZ}; for this reason, to make the presentation flow smoothly,  we postpone it to \ref{sec:appendix}. Once the system of Boltzmann type equation is derived, it allows to have a precise description of the evolution of the population densities. Unfortunately, working directly on Boltzmann-type kinetic equation systems, which include nonlinear integral operators, is difficult. For this reason, simplified models are used that can be handled more easily, while retaining as many physically relevant properties as possible.

In this direction, it is important to remark that Boltzmann-type systems can be simplified by assuming that most interactions are grazing, namely that they induce on post-interaction variables modifications of order $\eps \ll 1$ on the pre-interaction variables, a modification that can be easily obtained by scaling the parameters and time by $\eps$ \cite{pareschi2003spectral,alexandre2004landau}. Classically, a grazing interaction of type \fer{eq:SK} is given by 
 \be\label{eq:SKE}
x^*_\eps =x-\eps \phi(y)x+\sqrt\eps \eta(y)x,
 \ee
 which is such that the deterministic part of the variation is scaled by a factor $\eps$, and the variance of the random contribution has the same order in $\eps$ of the deterministic part. This allows to maintain in the limit equations both the contributions of the deterministic and random parts. 

 Moreover, under scaling \fer{eq:SKE},  conditions \fer{eq:neg}--\fer{eq:small-var} reduce to the weaker conditions
 \be\label{eq:weak}
\-\sqrt{\eps} \eta(y) \le 1-\eps\beta, \quad \sigma < 2\beta -\eps\beta^2,
 \ee
 which clearly improve the positivity conditions on the post-interaction values.
 
 Likewise, the linear interaction \fer{eq:redi1} is modified by substituting $\gamma$ with $\eps\gamma$. Even in this case, the scaling relaxes the condition that ensures the positivity of the post-interaction state, which now reads $\eps\gamma\theta <1$. 

 Note that, thanks to \fer{eq:weak} and to the relaxed condition on the linear parameter $\gamma$, the mean values $m_{K,\eps}(t),K\in \mathcal C$ remain strictly positive in time for any value of $\eps\ll 1$.
 
 Lastly, to observe an effective variation in the time of grazing interactions, the time $t$ has to be scaled according to $t/\eps$ \cite{MTZ}.
 
 The importance of the grazing regime is closely related to the possibility to change the integro-differential Boltzmann-type system with a system of coupled Fokker--Planck type equations, which can more easily treated, while maintaining most of the physical characteristics of the Boltzmann one \cite{MTZ}. Leaving details to the \ref{sec:appendix}, in the grazing regime the kinetic system describing the evolution of the five densities $f_K(x,t)$, $K \in \mathcal C$ consists of a system of five coupled Fokker--Planck equation with time dependent coefficients of diffusion and linear in $x$ drift of the form
\be\label{eq:FP-graz}
\frac{\partial f_K(x,t)}{\partial t}=\frac \partial{\partial x}\left[\frac{\sigma_K(t)}2\frac{\partial}{\partial x}\left(x^2f_K(x,t)\right)+ b_K(x,t)f_K(x,t)\right].
\ee
In \fer{eq:FP-graz}, for any given $K\in \mathcal C$, $\sigma_K(t) >0$ is the time-dependent coefficients of diffusion, while
\be\label{eq:dri}
b_K(x,t) =\lambda_K(t) x - \mu_K(t),
\ee 
with $\lambda_K(t),\mu_K(t)>0$ is the linear drift.
The Fokker--Planck type equations are usually coupled with standard no-flux boundary conditions, which read
\be\label{eq:bc-mass}
 \left. \sigma_k(t) \frac \partial{\partial x}[x^2 f_K(x,t)] +b_K(x,t) \right|_{x=0} =0, \qquad t >0
\ee
and 
\be\label{eq:bc-mean}
 \left. \sigma_k(t) x^2 f_K(x,t)\right|_{x=0} =0, \qquad t >0.
\ee
The vanishing of the boundary term at infinity follows by choosing initial data with a smooth and rapid decay. For an exhaustive discussion on the physical meaning of these conditions we refer to \cite{Furioli} (cf. also \cite{bondesan2025lotka,MTZ}). 



The five diffusion coefficients $\sigma_K(t), K\in \mathcal C$ in \fer{eq:FP-graz} are given by
\begin{equation}\label{eq:diffu}
    \begin{aligned}
        &\sigma_S(t) ={\sigma_{SI}}m_I(t)+{\sigma_{SJ}}m_J(t)+{\sigma_{SR}}m_R(t),\qquad
        \sigma_I(t)={\sigma_{I}}m_S(t)\\
        &\sigma_J(t)= {\sigma_{J}}m_S(t)+{\sigma_{RJ}}m_S(t) \qquad
        \sigma_T ={\sigma_{TS}}, \qquad
        \sigma_R={\sigma_{RS}}.
    \end{aligned}
\end{equation}
Clearly, these coefficients take into account the different random parts of the interactions which move individuals from one compartment to the other. What is important to outline is that binary interactions give rise to diffusion coefficients which take into account both populations taking part to interactions,  one or both of them in the form of their time-dependent mean values, which, in accord with the previous discussion, are positive. In contrast, linear interactions are defined by  constant in time coefficients of diffusion. Likewise, for $K\in \mathcal C$ the linear drift terms 
in \fer{eq:FP-graz} are given by
\begin{equation}\label{eq:drift}
    \begin{aligned}
        &b_S(x,t) = (\beta_Im_I(t)+\beta_Jm_J(t)+\beta_R m_R(t))x-(\alpha_Tm_T(t)+\alpha_R m_R(t))\\
        &b_I(x,t) =(\gamma_T\theta_T+\gamma_{IR}\theta_{IR}-\beta_Im_S(t))x-(\gamma_T(\theta_T-1)+\gamma_{IR}(\theta_{IR}-1))m_I(t)-\gamma_{JI}m_J(t)\\
        &b_J(x,t) = (\gamma_{JI}\theta_{JI}+\gamma_{R}\theta_{R}-\beta_Jm_S(t))x-(\gamma_{JI}(\theta_{JI}-1)+\gamma_{R}(\theta_{R}-1))m_J(t)-\beta_{R}m_S(t)m_R(t)\\
        &b_T(x,t)= \alpha_Tx-\gamma_{T}m_I(t)\\
        &b_R(x,t) =\alpha_Rx-\left(\gamma_{IR}m_I(t)+\gamma_Rm_J(t)\right)
    \end{aligned}
\end{equation}
It is important to note that, due to the presence of negative terms into the coefficients of $x$, the drift operators corresponding to both the populations of infected people are pure drift operators only if the mean value of the susceptible  population is suitably bounded from above. 

\subsection{Moment systems}\label{sec:moments}

Thanks to the particular structure of the Fokker--Planck equation of type \fer{eq:FP-graz}, where the values of the diffusion coefficients and drift are given in \fer{eq:diffu} and \fer{eq:drift}, we can compute in a standard way the evolution in time of the principal moments of the probability density solution $f_K(x,t)$. In particular, since the diffusion term of the Fokker--Planck equation \fer{eq:FP-graz} does not contribute to  the evolution of the mean value of the density $f_K(x,t)$,  this evolution is governed by the drift term alone. Taking into account  \fer{eq:drift} one can easily conclude that the positive mean values $m_K(t), K\in \mathcal C$ of the populations satisfy the system 
\begin{equation}\label{eqmean}
\begin{aligned}
    \frac{d m_S(t)}{dt}&=-\beta_Im_S(t)m_I(t)-\beta_Jm_S(t)m_J(t)-\beta_Rm_S(t)m_R(t)+\alpha_Tm_T(t)+\alpha_Rm_R(t)\\
    \frac{d m_I(t)}{dt}&=\beta_I m_S(t)m_I(t)+\gamma_{JI}m_J(t)-\gamma_T m_I(t)-\gamma_{IR}m_I(t)\\
    \frac{d m_J(t)}{dt}&=\beta_Jm_S(t)m_J(t)+\beta_Rm_S(t)m_R(t)-\gamma_{JI}m_J(t)-\gamma_Rm_J(t)\\
    \frac{d m_T(t)}{dt}&=\gamma_T m_I(t)-\alpha_T m_T(t)\\
    \frac{d m_R(t)}{dt}&=\gamma_{IR}m_I(t)+\gamma_Rm_J(t)-\alpha_R m_R(t).
\end{aligned}
\end{equation}
The above system \eqref{eqmean} consists of five nonlinear ordinary differential equations, that is solved once a set of positive initial data $(m_S(0),\,m_I(0),\,m_J(0),\,m_T(0),\,m_R(0))\in \R_+^5$ is prescribed. As one can verify, system \fer{eqmean} coincides with the  one derived from the Boltzmann-based kinetic system \eqref{eqweak1}-\eqref{eqweak5}. Let us note once more  that, in consequence of the kinetic description in terms of Boltzmann type equations, as shown in  \ref{sec:appendix}, the solutions of system \fer{eqmean} are positive, so that the total mass is preserved in time. Hence, without loss of generality we will assume in the following that 
\be\label{eq:mass1}
m_S(0)+m_I(0)+m_J(0)+m_T(0)+m_R(0)=1
\ee

System \fer{eqmean} describes the evolution in time of the mean values of individuals in the five compartments, and provides the classical picture of epidemic modeling based on a description in terms of ordinary differential equations \cite{hethcote2000}. Clearly, positivity of the solutions to system \fer{eqmean} can be proven directly without resorting to the Boltzmann picture.

 Unlike the classical approach, the knowledge of the time evolution of the probability densities in the compartments allows us to compute other important mean quantities, the main of which are the variances $v_K(t), K \in \mathcal C$, defined in \fer{eqvar2}. 
To this extent, it is enough to compute the second moment of the solution $f_K(x,t)$ to the Fokker--Planck equation \fer{eq:FP-graz}, and to apply \eqref{eqvar2}. At the end, on obtains the system
\begin{equation}\label{eqvar}
    \begin{aligned}
        &\frac{dv_S(t)}{dt}=-\left[(2\beta_I-\sigma_{SI}) m_I(t)+(2\beta_J-\sigma_{SJ})m_J(t)+(2\beta_R-\sigma_{SR})m_R(t)\right]v_S(t)\\
        &\hspace{1.5 cm}+\left(\sigma_{SI}m_I(t)+\sigma_{SJ}m_J(t)+\sigma_{SR}m_R(t)\right)m_S^2(t)\\
        &\frac{dv_I(t)}{dt}=-\left[2\gamma_T\theta_T+2\gamma_{IR}\theta_{IR}-\sigma_Im_S(t)-2\beta_Im_S(t)\right]v_I(t)+\sigma_Im_S(t)m_I^2(t)\\
        &\frac{dv_J(t)}{dt}=-\left[2\gamma_{JI}\theta_{JI}+ 2\gamma_{R}\theta_{R}-\sigma_Jm_S(t)-\sigma_{RJ}m_S(t))-2\beta_Jm_S(t)\right]v_J(t)\\
        &\hspace{1.5cm}+(\sigma_J+\sigma_{RJ})m_S(t)m_J^2(t)\\
        &\frac{dv_T(t)}{dt}=-(2\alpha_T-\sigma_{TS})v_T(t)+\sigma_{TS}m_T^2(t)\\
        &\frac{dv_R(t)}{dt}=-(2\alpha_R-\sigma_{RS})v_R(t)+\sigma_{RS}m_R^2(t).
    \end{aligned}
\end{equation}
It is remarkable that, at variance with the simplified Fokker--Planck approach,  the evolution of variances in the kinetic equations \eqref{eqweak1}--\eqref{eqweak5}, does not take a closed form. 

Once system \eqref{eqmean} is solved,  \eqref{eqvar} is a system of five linear ordinary differential equations. Since the mean values satisfy the bounds $0< m_K(t) \le 1, K \in \mathcal C$, both positivity and  boundedness of the variances  follows provided the coefficients of the terms $v_K(t), K \in \mathcal C$ are negative. These conditions require suitable smallness assumptions on the various constants $\sigma_K$ and $\sigma_{KL}$, $K,L \in \mathcal C$ characterizing the random part of the microscopic interactions.
\begin{equations}\label{eq:cond-v}
& 2\beta_I > \sigma_{SI}, \quad 2\beta_J > \sigma_{SJ}, \quad 2\beta_R > \sigma_{SR}, \quad
 2\gamma_T\theta_T + 2\gamma_{IR}\theta_{IR}  > \sigma_I +2\beta_I, \\ & 2\gamma_{JI}\theta_{JI} + 2\gamma_R\theta_R > \sigma_J +\sigma_{RJ} + 2\beta_J, \quad 2\alpha_T > \sigma_{TS}, \quad 2\alpha_R > \sigma_{RS}. 
\end{equations}

\subsection{Equilibria of moment systems}\label{sec:equi-mean}
Before studying the temporal evolution of the densities $f_K(x,t), K \in \mathcal C$, solutions of the Fokker--Planck equations of type \fer{eq:FP-graz}, we are interested in proving that the system of ordinary differential equations  \eqref{eqmean}--\eqref{eqvar} describing the evolution of the mean values and variances in the AMR model has or not a (unique) equilibrium point, and if the solutions converge towards this point for large times. 

For possible and feasible equilibria of system \fer{eqmean} we will use the notation
\[
(m_S^{\infty},\,m_I^{\infty},\,m_J^{\infty},\,m_T^{\infty},\,m_R^{\infty}),
\] 
which, in view of condition \fer{eq:mass1} is such that $\sum_{K\in \mathcal C} m_K^\infty =1$.
Clearly, the existence of a stationary point for system \fer{eqmean}  needs to be coupled with clear compatibility conditions on the various parameters which characterize the passage from one compartment to another. To this aim, it is of primary importance that these parameters are chosen to guarantee a physically consistent equilibrium. 
In what follows we identify two main constraints, who involve the maximal rates of infections of the susceptible individuals. These constraints, which read
\be\label{eq:constraint}
\beta_I > \gamma_T + \gamma_{IR};\qquad \beta_J > \gamma_R + \gamma_{JI}
\ee
are obtained from the simple assumption that the mean numbers of infected individuals which move towards another compartment can not be greater than the mean number of infected coming from the compartment of susceptible.

If conditions \fer{eq:constraint} are verified, the stationary value of susceptible individuals, which equals
\be\label{eq:s-inf}
m_S^\infty = \frac{\alpha_Tm_T^\infty+ \alpha_Rm_R^\infty}{\beta_I m_I^\infty+ \beta_Jm_J^\infty+ \beta_Rm_R^\infty},
\ee
is such that $m_S^\infty <1$. This can be easily verified by owing to 
the fourth and fifth equations, which at equilibrium  give the relations 
\begin{equations}\label{eq:4-5}
    m_T^{\infty}&=\frac{\gamma_T}{\alpha_T}m_I^{\infty}\\
    m_R^{\infty}&=\frac{\gamma_{IR}}{\alpha_R}m_I^{\infty}+\frac{\gamma_R}{\alpha_R}m_J^{\infty}.
\end{equations}
Indeed, substituting the values of $m_T^\infty$ and $m_R^\infty$, as given by \fer{eq:4-5} on the numerator of \fer{eq:s-inf} shows the result.

Furthermore, using  relations \fer{eq:4-5} into the third equation of \fer{eqmean} we obtain
\begin{equation}\label{equil2}
\begin{aligned}
    m_S^{\infty}&=\frac{(\gamma_{JI}+\gamma_R)m_J^{\infty}}{\beta_J m_J^{\infty}+\beta_Rm_R^{\infty}}
    =\frac{(\gamma_{JI}+\gamma_R)m_J^{\infty}}{(\beta_J+\frac{\beta_R\gamma_R}{\alpha_R})m_J^{\infty}+\beta_R\frac{\gamma_{IR}}{\alpha_R}m_I^{\infty}}\\
    &=\frac{\alpha_R(\gamma_{JI}+\gamma_R)m_J^{\infty}}{(\alpha_R\beta_J+\beta_R\gamma_R)m_J^{\infty}+\beta_R\gamma_{IR}m_I^{\infty}}.
\end{aligned}
\end{equation}
Likewise, from the second equation in \fer{eqmean} we get
\begin{equation}\label{equil3}
    m_S^{\infty}=\frac{(\gamma_T+\gamma_{IR})m_I^{\infty}-\gamma_{JI}m_J^{\infty}}{\beta_Im_I^{\infty}}.
\end{equation}
Equating expressions \eqref{equil2} and \eqref{equil3} implies a relationship between $m^\infty_I$ and $m^\infty_J$, given by
\[
    \frac{\alpha_R(\gamma_{JI}+\gamma_R)m_J^{\infty}}{(\alpha_R\beta_J+\beta_R\gamma_R)m_J^{\infty}+\beta_R\gamma_{IR}m_I^{\infty}}=\frac{(\gamma_T+\gamma_{IR})m_I^{\infty}-\gamma_{JI}m_J^{\infty}}{\beta_Im_I^{\infty}},
\]
that can be rewritten as
\begin{equation}\label{equil5}
    \beta_I\alpha_R(\gamma_{JI}+\gamma_R)m_J^{\infty}m_I^{\infty}=\left((\gamma_T+\gamma_{IR})m_I^{\infty}-\gamma_{JI}m_J^{\infty}\right)\left((\alpha_R\beta_J+\beta_R\gamma_R)m_J^{\infty}+\beta_R\gamma_{IR}m_I^{\infty}\right).
\end{equation}
To simplify notations, let us define positive constants $A,B,C$ by
\begin{equation}
        A=\gamma_R + \gamma_{JI}; \quad
        B=\gamma_T+\gamma_{IR};\quad
        C=\alpha_R\beta_J+\gamma_R\beta_R.
\end{equation} 
Straightforward computations then show that \eqref{equil5} rewrites
\begin{equation}\label{equil6}
    C\gamma_{JI}(m_J^{\infty})^2+\left(A\beta_I\alpha_R+\beta_R\gamma_{JI}\gamma_{IR}-BC\right)m_I^{\infty}m_J^{\infty}-B\beta_R\gamma_{IR}(m_I^{\infty})^2=0.
\end{equation}
We introduce the ratio
$r={m_J^{\infty}}/{m_I^{\infty}}$.
In terms of $r$,  \eqref{equil6} takes the form of the second-order equation
\begin{equation}\label{equil7}
    C\gamma_{JI}r^2+\left(A\beta_I\alpha_R+\beta_R\gamma_{JI}\gamma_{IR}-BC\right)r-B\beta_R\gamma_{IR}=0.
\end{equation}
Let $D$ by defined as
$$D=A\beta_I\alpha_R+\beta_R\gamma_{JI}\gamma_{IR}-BC.$$
Equation \eqref{equil7} has a unique positive solution given by
\begin{equation}\label{equil8}
   \bar r=\frac{-D+\sqrt{D^2+4BC\beta_R\gamma_{JI}\gamma_{IR}}}{2C\gamma_{JI}}.
\end{equation}
Now, since $m_J^{\infty}=\bar rm_I^{\infty}$, using this relation into \fer{eq:4-5}  gives
\begin{equation}\label{equil9}
        m_J^{\infty}=\bar r m_I^{\infty}; \quad
        m_T^{\infty}=\frac{\gamma_T}{\alpha_T}m_I^{\infty};\quad
        m_R^{\infty}=\frac{\gamma_{IR}+\bar r\gamma_R}{\alpha_R}m_I^{\infty}.
\end{equation}
Moreover,  from \fer{equil2} we get
\be\label{equyil10}
m_S^{\infty}=\frac{\alpha_R(\gamma_{JI}+\gamma_R)\bar r}{(\alpha_R\beta_J+\beta_R\gamma_R)\bar r +\beta_R\gamma_{IR}} < 1,
\ee
in view of the constraints \fer{eq:constraint}. 
By mass conservation \fer{eq:mass1}, and by \fer{equil9}, we have 
\[
m_I^\infty= 1-m_S^\infty -\bar r m_I^{\infty}
        -\frac{\gamma_T}{\alpha_T}m_I^{\infty}-
        \frac{\gamma_{IR}+\bar r\gamma_R}{\alpha_R}m_I^{\infty},
\]
from which, since $m_S^\infty <1$, we can compute the unique positive value of $m_I^\infty$, given by
\be\label{eq:m_I}
m_I^\infty = \frac{\alpha_T\alpha_R(1-m_S^\infty )}{\alpha_T\alpha_R(1+\bar r )
        +\alpha_R{\gamma_T}+
        \alpha_T({\gamma_{IR}+\bar r\gamma_R})}.
\ee
The remaining values at equilibrium are obtained from \fer{equil9} by substituting the value of $m_I^\infty$ defined in \fer{eq:m_I}. 

{ We finally have

\begin{equation}\label{expliciteq}
    \begin{aligned}
    m_S^{\infty}&=\frac{\alpha_R(\gamma_{JI}+\gamma_R)\bar r}{(\alpha_R\beta_J+\beta_R\gamma_R)\bar r +\beta_R\gamma_{IR}}\\
    m_I^{\infty}&=\frac{
\alpha_T \alpha_R
\left(
\left[\alpha_R(\beta_J - \gamma_{JI} - \gamma_R) + \beta_R \gamma_R \right]\bar r
+ \beta_R \gamma_{IR}
\right)
}{
\left[(\alpha_R \beta_J + \beta_R \gamma_R)\bar r + \beta_R \gamma_{IR}\right]
\left[
\alpha_T \alpha_R (1+\bar r)
+ \alpha_R \gamma_T
+ \alpha_T (\gamma_{IR} + \bar r \gamma_R)
\right]
}\\
    m_J^{\infty}&=\frac{
\bar{r}\alpha_T \alpha_R
\left(
\left[\alpha_R(\beta_J - \gamma_{JI} - \gamma_R) + \beta_R \gamma_R \right]\bar r
+ \beta_R \gamma_{IR}
\right)
}{
\left[(\alpha_R \beta_J + \beta_R \gamma_R)\bar r + \beta_R \gamma_{IR}\right]
\left[
\alpha_T \alpha_R (1+\bar r)
+ \alpha_R \gamma_T
+ \alpha_T (\gamma_{IR} + \bar r \gamma_R)
\right]
}\\
    m_T^{\infty}&=\frac{
\gamma_T \alpha_R
\left(
\left[\alpha_R(\beta_J - \gamma_{JI} - \gamma_R) + \beta_R \gamma_R \right]\bar r
+ \beta_R \gamma_{IR}
\right)
}{
\left[(\alpha_R \beta_J + \beta_R \gamma_R)\bar r + \beta_R \gamma_{IR}\right]
\left[
\alpha_T \alpha_R (1+\bar r)
+ \alpha_R \gamma_T
+ \alpha_T (\gamma_{IR} + \bar r \gamma_R)
\right]
}\\
    m_R^{\infty}&=\frac{
\alpha_T (\gamma_{IR}+\bar{r}\gamma_R)
\left(
\left[\alpha_R(\beta_J - \gamma_{JI} - \gamma_R) + \beta_R \gamma_R \right]\bar r
+ \beta_R \gamma_{IR}
\right)
}{
\left[(\alpha_R \beta_J + \beta_R \gamma_R)\bar r + \beta_R \gamma_{IR}\right]
\left[
\alpha_T \alpha_R (1+\bar r)
+ \alpha_R \gamma_T
+ \alpha_T (\gamma_{IR} + \bar r \gamma_R)
\right].
}\\
    \end{aligned}
\end{equation}
}
In order to further strengthen the results, we need to understand whether the solution $\mathbf{m}(t)$ of system \eqref{eqmean} converges to the equilibrium $\mathbf{m}^{\infty}$ derived above. However, obtaining this result is not straightforward because of the large number of parameters involved and the multiple interactions present in the system. Due to the strongly coupled nonlinear structure of system \eqref{eqmean}, a direct estimate of the distance between solutions (for instance in the $L^1$ norm as in \cite{bondesan2025lotka}) does not lead to a closed dissipative inequality. For this reason, the stability analysis of the equilibrium is performed through the Jacobian matrix and the Routh--Hurwitz criterion, which provides necessary and sufficient conditions for local asymptotic stability, as shown in \ref{appendix:stability}.

\noindent Owing to the unique equilibrium of mean values, we easily get the unique asymptotic values of the variances, which, as previously remarked,  are positive and bounded in presence of  assumptions \fer{eq:cond-v} on the coefficients.
In this case, the asymptotic set of variances 
$(v_S^{\infty}, v_I^{\infty},v_J^{\infty},v_T^{\infty},v_R^{\infty})$  is given by
\begin{equation}\label{varequil}
    \begin{aligned}
v_S^{\infty}&=\frac{\left(\sigma_{SI}m_I^{\infty}+\sigma_{SJ}m_I^{\infty}+\sigma_{SR}m_I^{\infty}\right)(m_S^{\infty})^2}{(2\beta_I-\sigma_{SI})m_I^{\infty}+ (2\beta_J-\sigma_{SJ})m_J^{\infty}+(2\beta_R-\sigma_{SR})m_R^{\infty}}\\
        v_I^{\infty}&=\frac{\sigma_Im_S^{\infty}(m_I^{\infty})^2}{2\gamma_T\theta_T+2\gamma_{IR}\theta_{IR}-(\sigma_I+2\beta_I)m_S^{\infty}}\\
        v_J^{\infty}&=\frac{(\sigma_Jm_S^{\infty}+\sigma_{RJ}m_S^{\infty})(m_J^{\infty})^2}{2\gamma_{JI}\theta_{JI}+2\gamma_{R}\theta_{R}-(\sigma_J+2\beta_J +\sigma_{RJ})m_S^{\infty}}\\
        v_T^{\infty}&=\frac{\sigma_{TS}(m_T^{\infty})^2}{2\alpha_T-\sigma_{TS}}\\
        v_R^{\infty}&=\frac{\sigma_{RS}(m_R^{\infty})^2}{2\alpha_R-\sigma_{RS}}.
    \end{aligned}
\end{equation}

\subsection{A meanfield model for AMR}\label{sec:equi}

The discussion of Section \ref{sec:moments} allows to conclude that both the diffusion and the drift terms of the Fokker--Planck equation \fer{eq:FP-graz} satisfy the conditions of
Proposition 2 of Section 6 in \cite{bris2008existence}. This Proposition refers to a Fokker--Planck equation of divergence form, which in our case is equivalent to
\be\label{eq:FP-div}
\frac{\partial f_K(x,t)}{\partial t}=\frac \partial{\partial x}\left[\frac 12{\sigma_K^2(x,t)}\frac{\partial f_K(x,t)}{\partial x}+  b_K^*(x,t)f_K(x,t)\right],
\ee
where
\be\label{eq:FP-coe}
\sigma_K^2(x,t) = \sigma_K(t) x^2, \quad b_K^*(x,t) = b_K(x,t)+ \sigma_K(t)x.
\ee
Moreover, in Proposition 2 of \cite{bris2008existence}, the variable $x \in \R$. 
The validity of Proposition 2 follows from the fact that $\sqrt{\sigma_K(x,t)}$ and $b_K^*(x,t)$ are linear in $x$, while $\sigma_K(t), \lambda_K(t)$ and $\mu_K(t)$, in reason of their dependence from the mean values $m_K(t)$, belong to $L_1(0,T)\cap L_\infty(0,T)$ for any $T >0$.

Furthermore, as proven in Section 2.3 of Ref. \cite{torregrossa2018} for equation \fer{eq:FP-div} with time-independent coefficients, if the initial value is different from zero only for $x \in \R_+$, the solution to  equation \fer{eq:FP-div} remains different from zero only for $x \in \R_+$, and this property can be directly extended to the case of time-dependent coefficients.

Under the aforementioned conditions,  Proposition 2 of \cite{bris2008existence} ensures that for any time $T>0$ and each initial datum in $L_1(\R_+) \cap L_\infty(\R_+)$, equation \fer{eq:FP-div}, and consequently the Fokker--Planck equation \fer{eq:FP-graz} has a unique solution in the space
\be\label{eq:glob}
f_K(x,t) \in L_\infty\left( [0,T], L_1(\R_+) \cap L_\infty(\R_+) \right), \quad 
\sqrt{\sigma_K(t)}\, x \frac{\partial f_K(x,t)}{\partial x} \in  L_2\left( [0,T], L_2(\R_+)  \right).
\ee
Therefore, we can conclude, under natural hypotheses on the set of initial values $f_{0,K}(x), K \in \mathcal C$, that the AMR system of Fokker--Planck equations has a unique global solution satisfying \fer{eq:glob}. Moreover, provided the diffusion constants satisfy \fer{eq:cond-v}, the probability densities $f_K(x,t),K \in \mathcal C$ have bounded variances. 

In addition to the knowledge of the behavior of moments of the solution densities, as given by systems \fer{eqmean} and \fer{eqvar}, it is usual to to get additional information on the behavior of the system by studying directly various quantities related to the solution itself. Among them, it has a big relevance to study the so-called local equilibria of the Fokker--Planck equations  \fer{eq:FP-graz}, namely the solution  of unit mass obtained from \fer{eq:FP-graz} by setting to zero the right-hand side. Hence, the local equilibrium density $f^q_K(x,t)$ is found by solving the first-order differential equation
\be\label{eq:FP-quasi}
\displaystyle \frac{\sigma_K(t)}2\frac{\partial}{\partial x}\left(x^2f_K(x,t)\right)+ b_K(x,t)f_K(x,t)=0,
\ee
Recalling the definition of the drift term $b_K=\lambda_K(t) x- \mu_K(t)$, it is immediate to show that, for $K\in \mathcal C$
\be\label{eq:quasi}
f^q_K(x,t) = \frac{\omega_K(t)^{\nu_K(t)}}{\Gamma(\nu_K(t))}\left(\frac{1}{x}\right)^{\nu_K(t)+1}\exp\left\{-\frac{\omega_K(t)}x\right\},
\ee
where the shape parameter $\nu_K(t)$ and the scale parameter $\omega_K(t)$ are given by
\be\label{eq:scale}
\nu_K(t) = 1+ \frac{2\lambda_K(t)}{\sigma_K(t)}, \qquad \omega_K(t) = \frac{2\mu_K(t)}{\sigma_K(t)}.
\ee
Thus, for any fixed value of time $t >0$, the local equilibria  are  inverse Gamma densities of  mean and variance 
\be\label{eq:mom}
m^q_K(t) = \frac{\mu_K(t)}{\lambda_K(t)}, \qquad v^q_K(t) = \left( \frac{\mu_K(t)}{\lambda_K(t)}\right)^2 \frac{\sigma_K(t)}{2\lambda_K(t)-\sigma_K(t)}.
\ee

Note that the variance is finite provided $2\lambda_K(t)>\sigma_K(t)$, namely when, as remarked before, the diffusion coefficients are suitably small.
It is further important to remark that, provided $\mu_K(t) \to \mu_\infty$, $\lambda_K(t) \to \lambda_\infty$ and $\sigma_K(t) \to \sigma_\infty$, 
\be\label{eq:lim}
m^q_K(t) \to m_K^\infty, \qquad v^q_K(t) \to v_K^\infty,
\ee
namely to the stationary mean values and variances of the solutions of \fer{eq:FP-graz} derived in Section \ref{sec:moments}. Consequently, if \fer{eq:lim} hold, the local equilibria furnish a reasonable approximation to the true solutions, at least for large times.

\section{Large-time behavior}\label{sec:large}

As shown in \cite{MTZ}, a further important role of the local equilibria is their relationship with the large-time convergence to a steady state of the population densities in the various compartments, once the convergence of their mean values and variances to limit values has been shown to hold.

Let us briefly recall this result, by addressing the interested reader for details to Section 5 of Ref. \cite{MTZ}.

For any given time $t >0$, and $K\in \mathcal C$, the local equilibrium of the Fokker--Planck equation \fer{eq:FP-graz}  is the unique density function $f_K^q(x,t)$  of unit mass satisfying \fer{eq:FP-quasi}.

Therefore, since the right-hand side of the (linear) Fokker--Planck equation \fer{eq:FP-graz}  vanishes when evaluated in correspondence to the local equilibrium density, we have the identity
\begin{equations}\label{eq:base}
&\dfrac{\partial [f_K(x,t)- f_K^q(x,t)]}{\partial t}= - \dfrac{\partial f_K^q(x,t)}{\partial t} +\\
& \dfrac{\sigma_K(t)}{2}\dfrac{\partial^2}{\partial x^2}\left\{x^{2}[f_K(x,t)- f_K^q(x,t) \right\}+\dfrac{\partial}{\partial x}\left\{b_K(x,t)[f_K(x,t)-f_K^q(x,t)]\right\}\,.
\end{equations}
The large-time behavior of the linear equation \fer{eq:base} can be treated by passing to the Fourier transform, where, as usual, we define the transform of a probability density $f(x), x \in \R_+$ by
\[
\fc(\xi) = \int_{\R_+}f(x)e^{-i\xi x}\, dx.
\]
This procedure has been first applied to a Fokker--Planck equation of type \fer{eq:FP-graz} with constant coefficients in Ref. \cite{torregrossa2018}, and extended to the case with time-dependent coefficients in \cite{MTZ}. There it was shown that for any given positive constant $p$ satisfying $1/2 <p < 1$, by passing to Fourier transform the $\dot H_{-p}$--Sobolev norm of the difference $\fc_K(\xi,t)- \fc_K^q(\xi,t)$, of wide use in Statistics under the name of Energy distance \cite{auricchio2026kinetic},  satisfies the inequality 
\begin{equations}\label{inizio}
&\dfrac d{dt}{\int_\R |\xi|^{-2p} |\fc_K(\xi,t)- \fc_K^q(\xi,t)|^2}\,d\xi \le  \int_\R  |\xi|^{-2p}\left|\widehat{ \dfrac{\partial f_K^q}{\partial t}}(\xi,t)\right||\fc_K(\xi,t)- \fc_K^q(\xi,t)|\, d\xi +\\
& -(2p-1)\left(\sigma_K(t) \frac{3-2p}4 + \lambda_K(t) \right) {\int_\R |\xi|^{-2p} |\fc_K(\xi,t)- \fc_K^q(\xi,t)|^2}\,d\xi.
\end{equations}
The positivity of $\sigma_K(t)$ and $\lambda_K(t)$, coupled with the fact that $p >\frac 12$ imply that, in the absence of the contribution of the term containing the time derivative of the local equilibrium, the Energy distance of the difference converges exponentially to zero. 
Thus, one has to examine the contribution of this term.
In \cite{MTZ}, using  that the local equilibrium in an inverse Gamma density, the following upper bound has been derived
\begin{equations}\label{finale}
&\int_{\R}  |\xi|^{-2p}\left|\widehat{ \dfrac{\partial f_K^q}{\partial t}}(\xi,t)\right||\fc_K(\xi,t)- \fc_K^q(\xi,t)| \, d\xi \le \\
&\epsilon_K(t) C_p \left[m_K(t) +m_K^q(t)\right]^{\frac{1}{3-2p}}\left( \int_{\R}  |\xi|^{-2p}|\fc_S(\xi,t)- \fc_S^q(\xi,t)|^2 \, d\xi \right)^{\frac{2-2p}{3-2p}},
\end{equations}
where $\epsilon_K(t) \to 0$, and $C_p$ is an explicitly computable constant.
Since both $m_K(t)$ and $m_K^q(t)$ are less than $1$, in \fer{finale}
\[
C_p \left[m_K(t) +m_K^q(t)\right]^{\frac{1}{3-2p}}\le 2^{\frac{1}{3-2p}}C_p = D_p.
\]
Hence, by setting
\[
z(t) = \int_{\R}  |\xi|^{-2p}|\fc_K(\xi,t)- \fc_K^q(\xi,t)|^2 \, d\xi, 
\]
 thanks to \fer{inizio} and to \fer{finale} we have that $z(t)$ satisfies the differential inequality
\[
\frac{dz(t)}{dt} \le -c_p(t)  z(t) + \epsilon_S(t) D_p z(t)^{\frac{2-2p}{3-2p}},
\]
where the value of  $c_p(t)$ can be recovered from \fer{inizio}.  Since $\epsilon_S(t) \to 0$ as $t \to + \infty$,  $z(t)$ converges to zero as well.

\section{Numerical tests}\label{sec:numerics}
In this section, we provide some numerical tests to illustrate the features of the designed model. First, we show the consistency of the model with the obtained quasi-equilibrium distributions by numerically approximate the evolution of \eqref{eq:FP-graz}, characterized by a drift $b_K(t)$ as in \eqref{eq:drift} and nonconstant diffusion $\sigma_K(t)$ as in \eqref{eq:diffu}, $K \in\{S,I,J,T,R\}$, by means of the structure reserving scheme for Fokker-Planck equations proposed in \cite{PareschiZanella2018}. These methods are globally second order and gain arbitrary accuracy for large times, together with preservation of positivity and the solution, and a consistent entropy dissipation. In all the subsequent tests, we considered a domain $[0,L]$, $L = 300$ discretized by means of $N = 8001$. The discretization of the considered time interval has a time step $\Delta t>0$ such that $\Delta t = \Delta x/2$ to comply with positivity preservation of semi-implicit time integration methods discussed in \cite{PareschiZanella2018}. At the boundary $x=0$ we always consider a no-flux boundary condition. 

\subsection{\textbf{Test 1}. Convergence toward the equilibrium distributions}
 
We show the evolution of the numerical solution to \eqref{eq:FP-graz} for which we obtained convergence toward the quasi-equilibrium profiles defined in \eqref{eq:FP-quasi}. In more detail, we will consider the initial distributions
\[
\begin{split}
f_S(x,0) &= 
\begin{cases}
10, & x\in[3.95,4.05],\\
0, & \text{otherwise},
\end{cases}
\qquad
f_I(x,0) = 
\begin{cases}
10, & x\in[0.95,1.05],\\
0, & \text{otherwise},
\end{cases}
\\[0.4em]
f_J(x,0) &= 
\begin{cases}
10, & x\in[0.95,1.05],\\
0, & \text{otherwise},
\end{cases}
\qquad
f_T(x,0)= 
\begin{cases}
10, & x\in[0.95,1.05],\\
0, & \text{otherwise},
\end{cases}
\\[0.4em]
f_R(x,0)&= 
\begin{cases}
10, & x\in[0.45,0.55],\\
0, & \text{otherwise}.
\end{cases}
\end{split}
\]
In the following test, we fix the parameters as listed in Table \ref{tab:parameters}. They are chosen such that the above assumptions \eqref{eq:constraint} of the model are satisfied. For simplicity, we assume uniformity across compartments of the loss of immunity rate and recovery rate across treated $T$ and resistant $R$ compartments
\[
\begin{split}
\alpha_T &= \alpha_R = \alpha >0, \qquad  \gamma_T=\gamma_R=\gamma.
\end{split}\]
Furthermore, we assume a fixed diffusion coefficients across all the compartments
\[
\sigma_{SI}=\sigma_{SJ}=\sigma_{SR}=\sigma_{I}=\sigma_J=\sigma_{RJ}=\sigma_{TS}=\sigma_{RS}, 
\]
Finally, we fix a constant of redistribution coefficient
\[
    \theta_T=\theta_R=\theta_{IR}=\theta_{JI}=\theta.
\]

 \begin{table}[htbp]
\centering
\caption{Baseline values of the model parameters used in the numerical simulations.}
\label{tab:parameters}
\renewcommand{\arraystretch}{1.12}
\begin{tabular}{|c|c|p{8.2cm}|}
\hline
\textbf{Parameter} & \textbf{Value} & \textbf{Meaning} \\
\hline
$\alpha$ & $\frac{1}{50}$ & Loss of immunity rate\\ \hline
$\beta_I$ & $0.18$ & Infection rate due to interactions with sensitive-strain pathogen \\ \hline
$\beta_J$ & $0.18$ & Infection rate due to interactions with resistant-strain pathogen \\ \hline
$\beta_R$ & $0.12$ & Infection rate due to interactions with resistant-treated individuals  \\ \hline
$\gamma$ & $\frac{1}{14}$ & Recovery rate \\ \hline
$\gamma_{IR}$ & $\frac{1}{14}$ & Transition rate due to bad-treatment \\ \hline
$\gamma_{JI}$ & $\frac{1}{14}$ & Transition rate due to well-treatment  \\ \hline
$\sigma$ & $0.01$ & Diffusion coefficient \\
\hline
$\theta$ & $2.5$ & Costant of redistribution for well-trated compartment\\ \hline
\end{tabular}
\end{table}

\begin{figure}
    \centering
    \includegraphics[width=70mm]{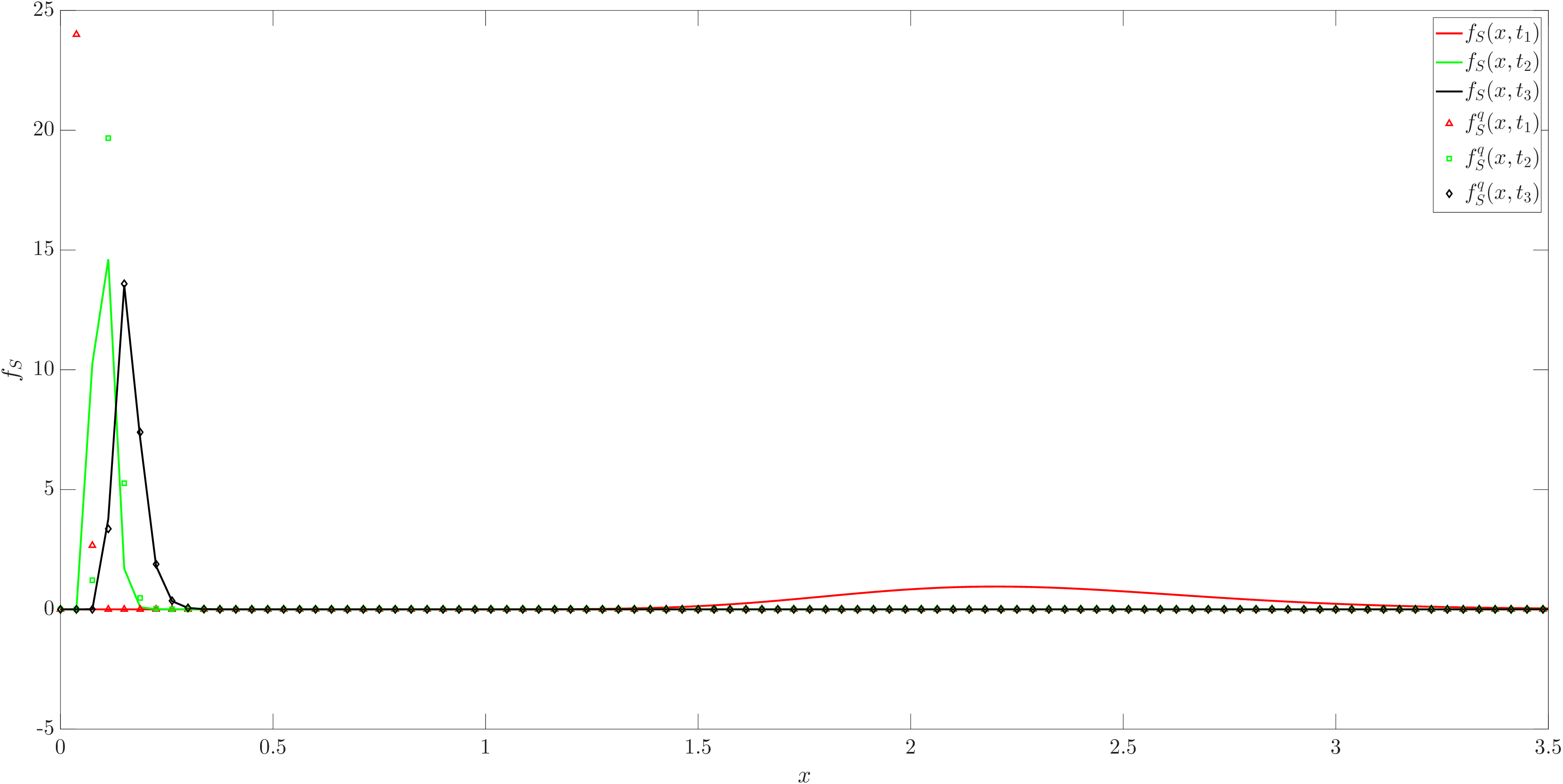}
    \includegraphics[width=70mm]{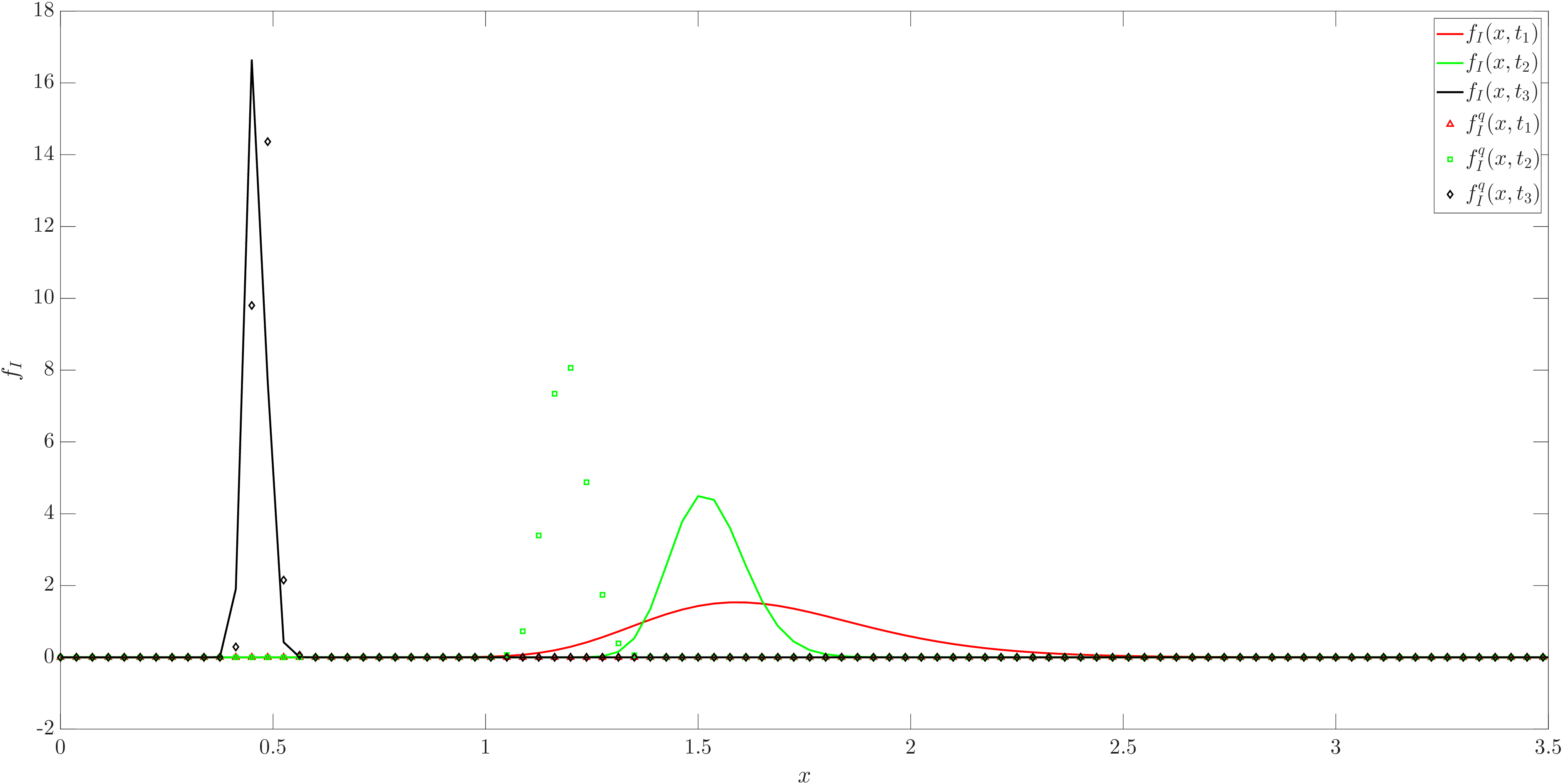}\\
    \includegraphics[width=70mm]{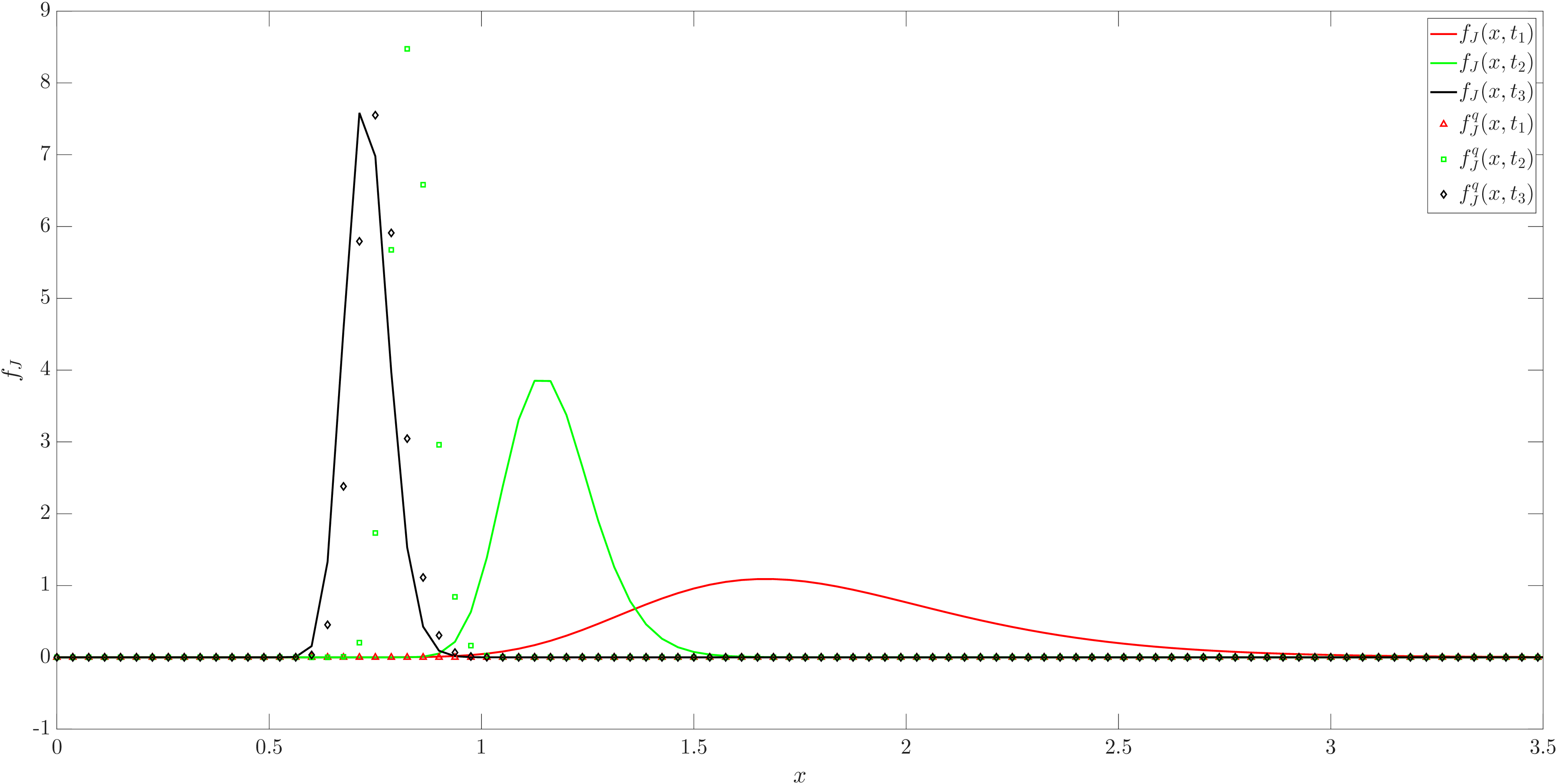}
    \includegraphics[width=70mm]{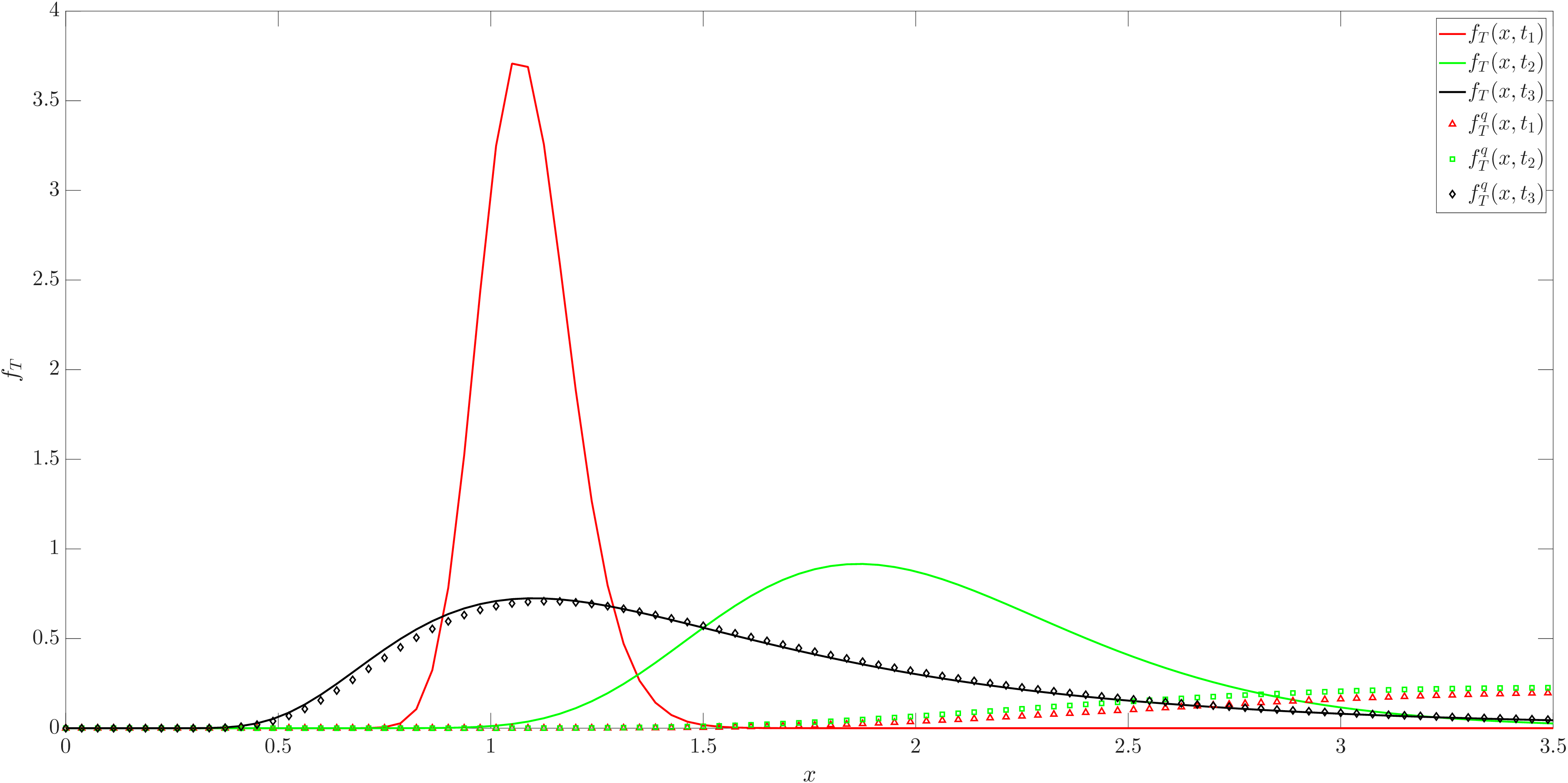}\\
    \includegraphics[width=70mm]{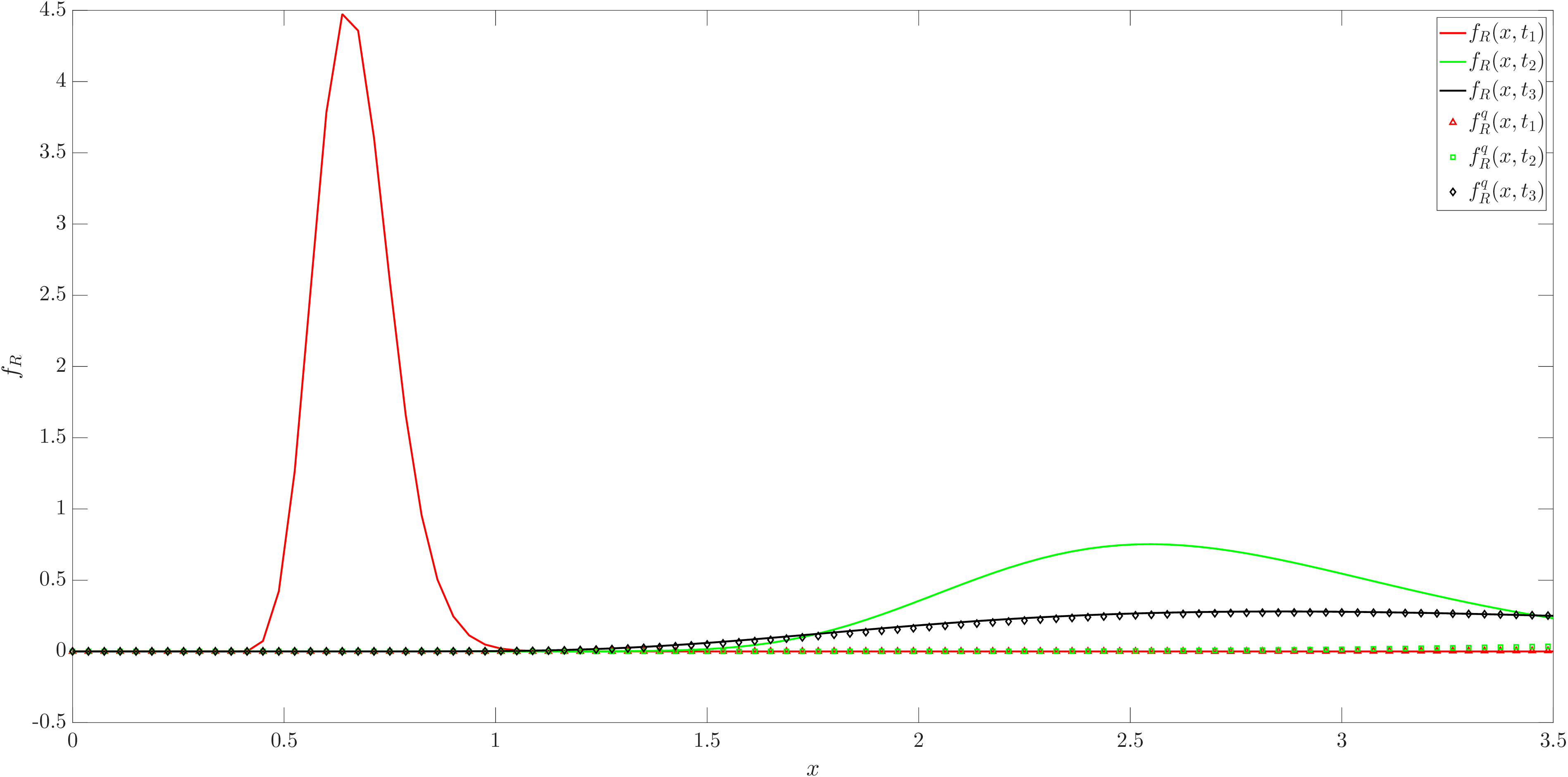}
    \caption{Evolution of the solution  $f_K(x,t)$ of the Fokker-Planck system and the related quasi-equilibrium $f^q_K(x,t)$, for each compartment of the system $K \in \{S,\,I,\,J,\,R,\,T\}$, for three different time-steps: $t=1$ (red), $t=10$ (green), and $t=300$ (black). We may observe how each distribution $f_K$ collapses toward its quasi-equilibrium $f_K^q$ for large times.}
    \label{qeevolution}
\end{figure}

\noindent The evolution of the densities $f_S(x,t)$, $f_I(x,t)$, $f_J(x,t)$, $f_T(x,t)$, $f_R(x,t)$ are approximated through a semi-implicit structure preserving scheme over the domain $[0,L]$, $L = 300$, discretized with $N = 8001$ gridpoints. The dynamics is integrated over the time horizon $[0,T]$, $T=300$ with time step $\Delta t = \Delta x/2$. In Figure \ref{qeevolution} we represent the approximations of the kinetic densities together with the quasi-equilibrium profiles $f_K^q$ defined in \eqref{eq:FP-quasi} at three time steps $t_1 = 1$, $t_2 = 10$, $t_3= 300$. We can observe how for large times the distributions $f_K(x,t)$ converge toward $f_K^q(x,t)$, therefore confirming the theoretical results in Section \ref{sec:large}.

\subsection{\textbf{Test 2}. Impact of the resistant strain}

\begin{figure}
    \centering
    \includegraphics[width=70 mm]{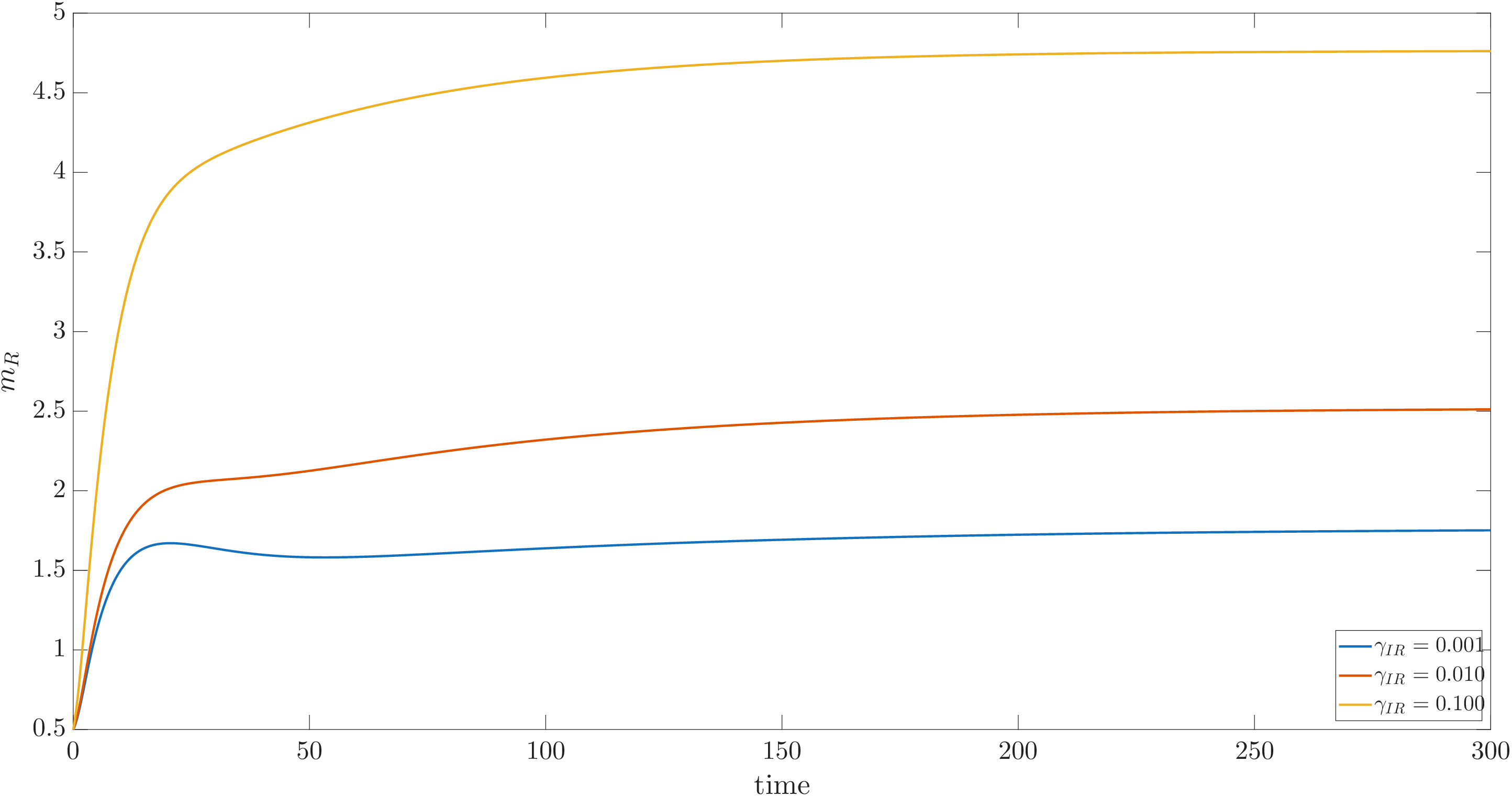}
    \includegraphics[width=70 mm]{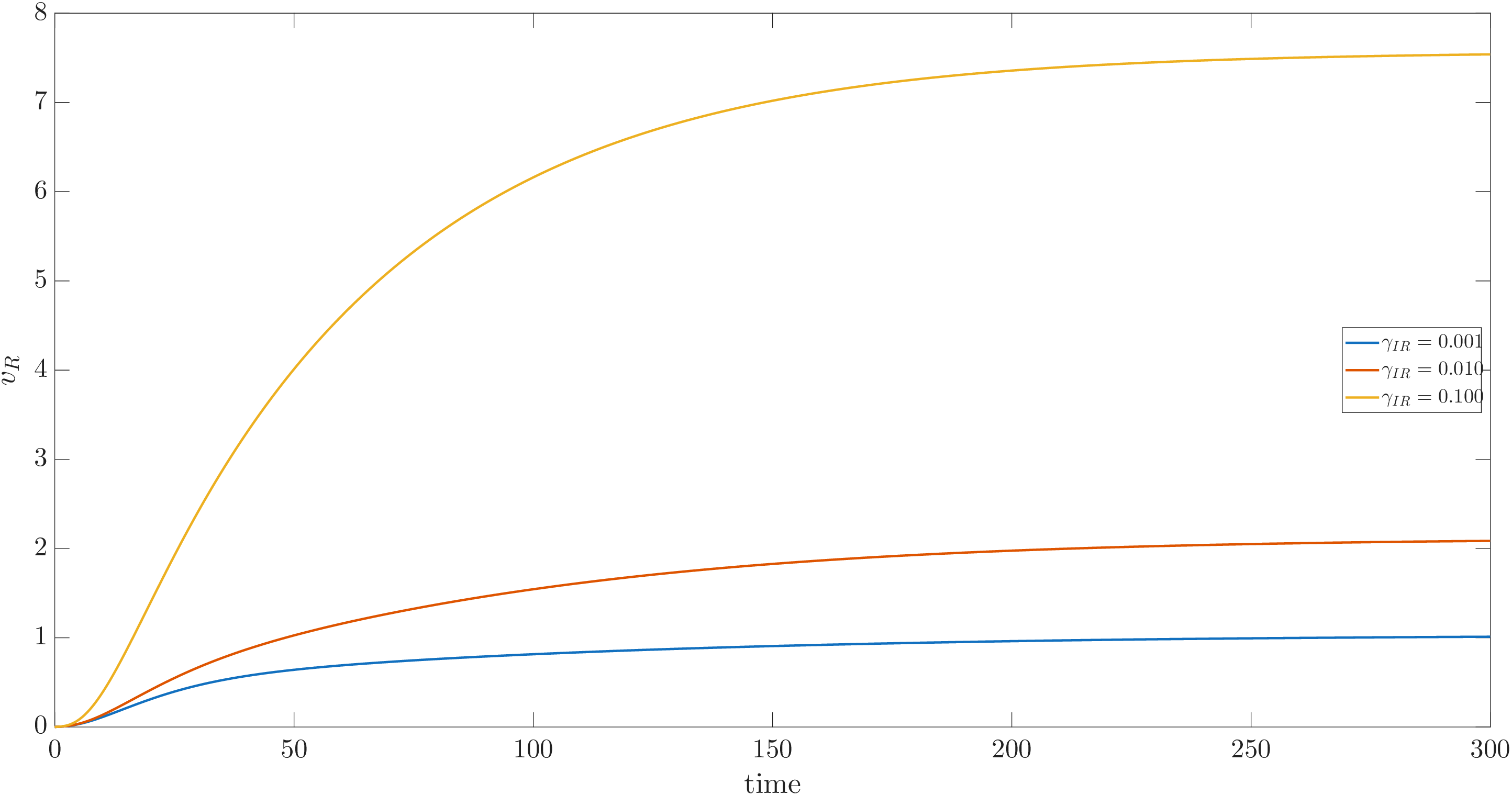}
    \includegraphics[width=70 mm]{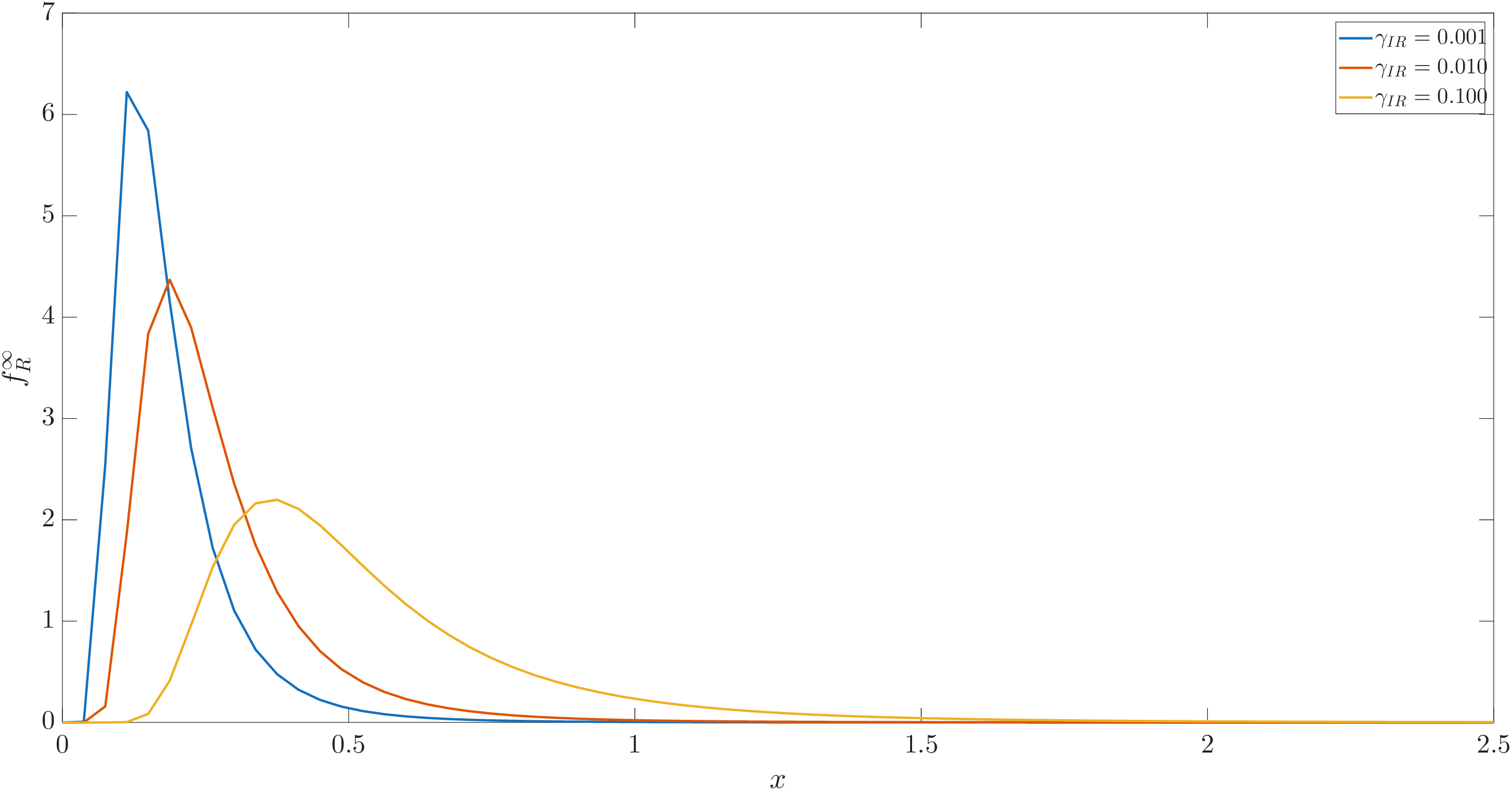}
    \includegraphics[width=70 mm]{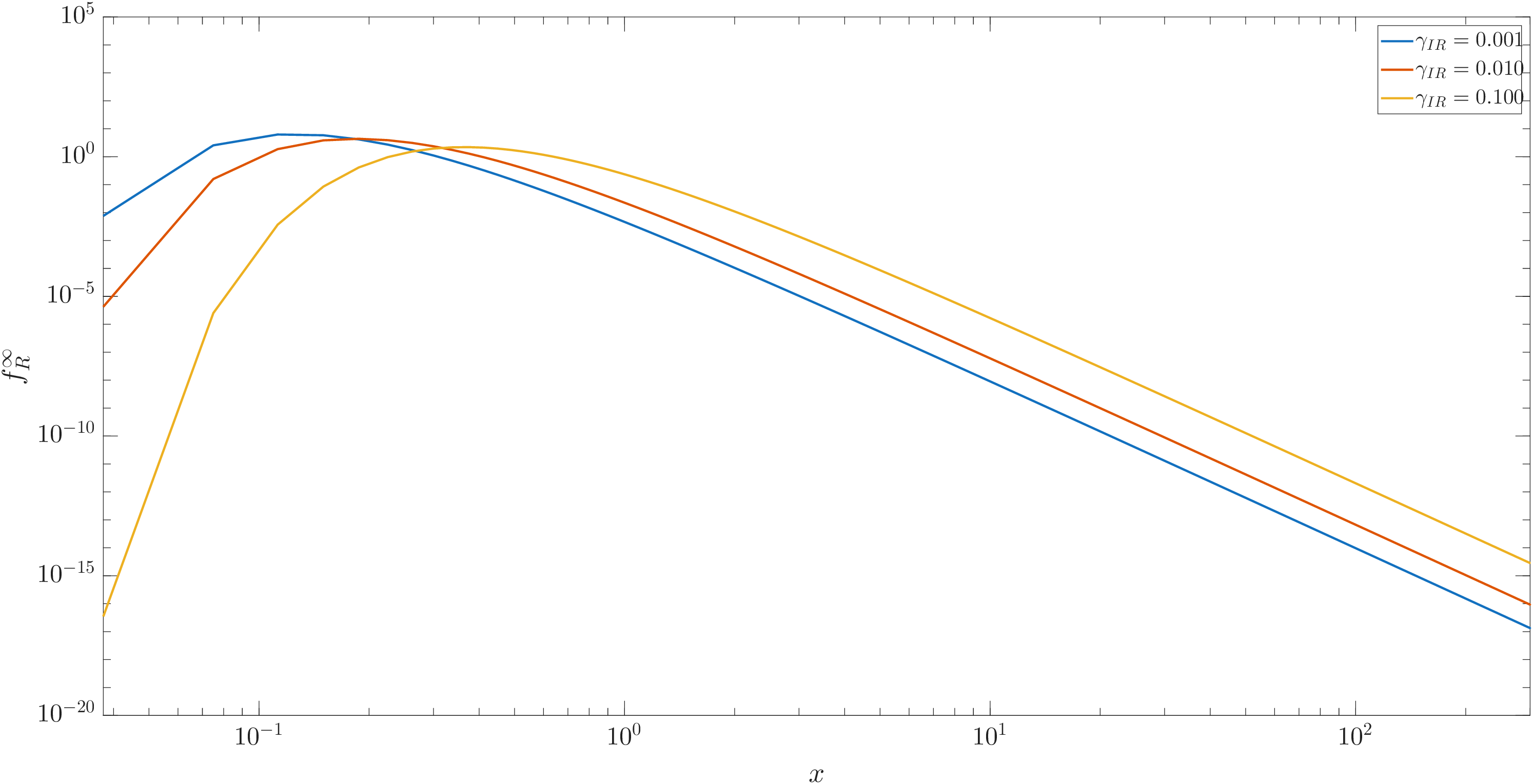}
    \caption{Different behaviors of the evolution for the resistant compartment, i.e. $R$, with respect to different values of the parameter $\gamma_{IR}$. From top to bottom and from left to right: the closed form of the mean $m_R$, the closed form of the variance $v_R$, and the equilibrium $f^{\infty}_R$, displayed using both linear and log–log scales.}
    \label{gammaIRpic}
\end{figure}

\begin{figure}
    \centering
    \includegraphics[width=70 mm]{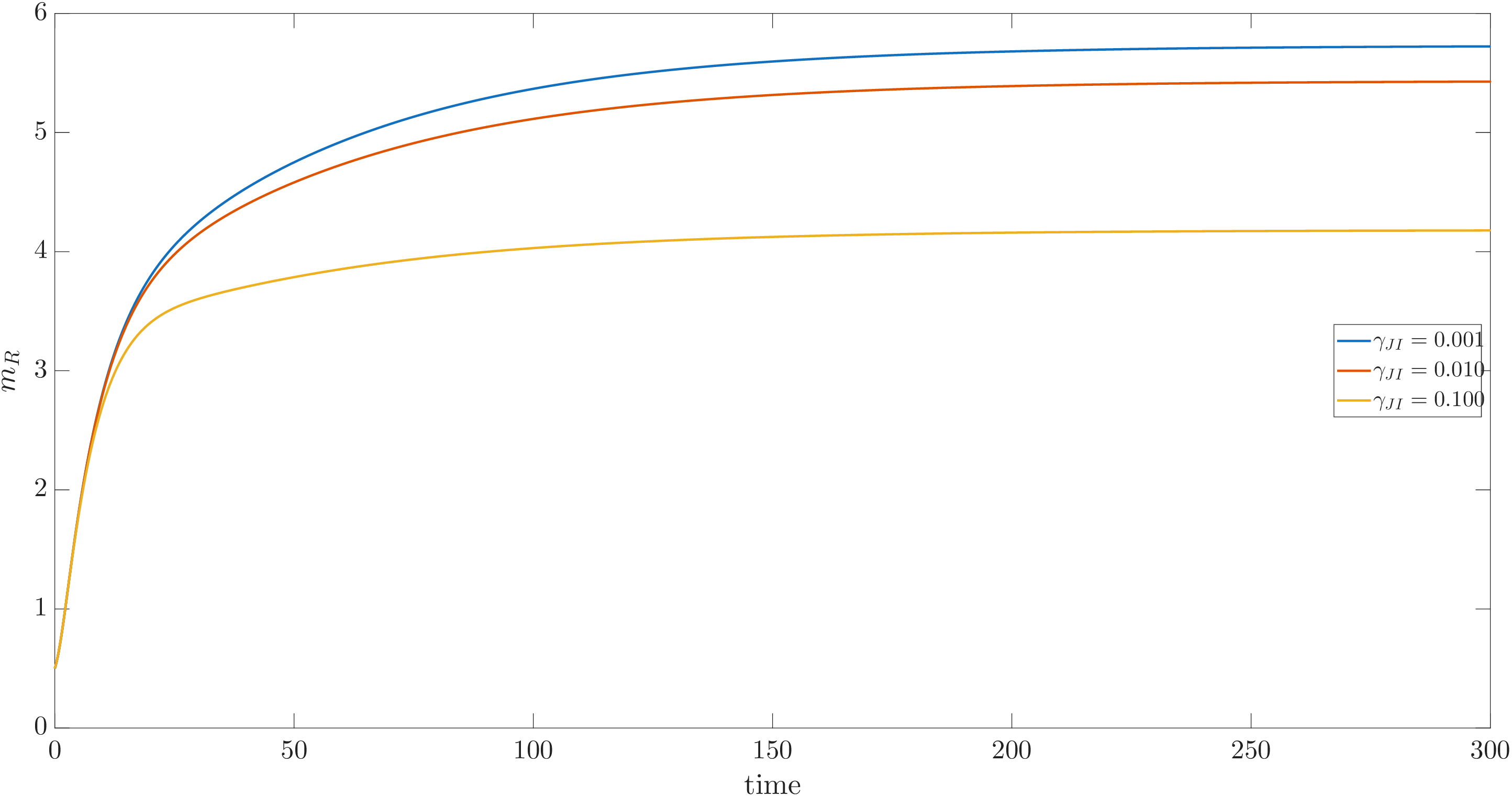}
    \includegraphics[width=70 mm]{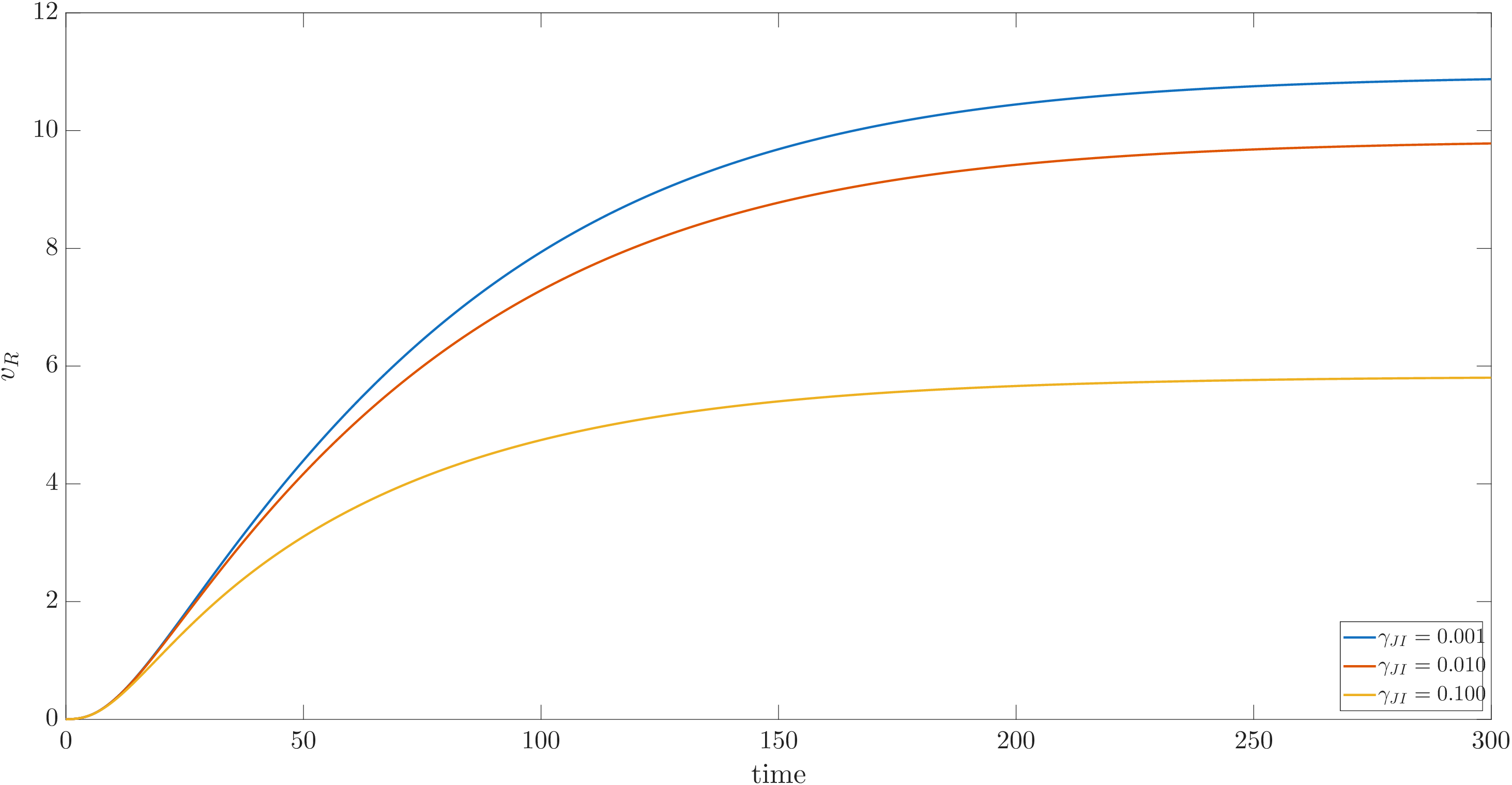}
    \includegraphics[width=70 mm]{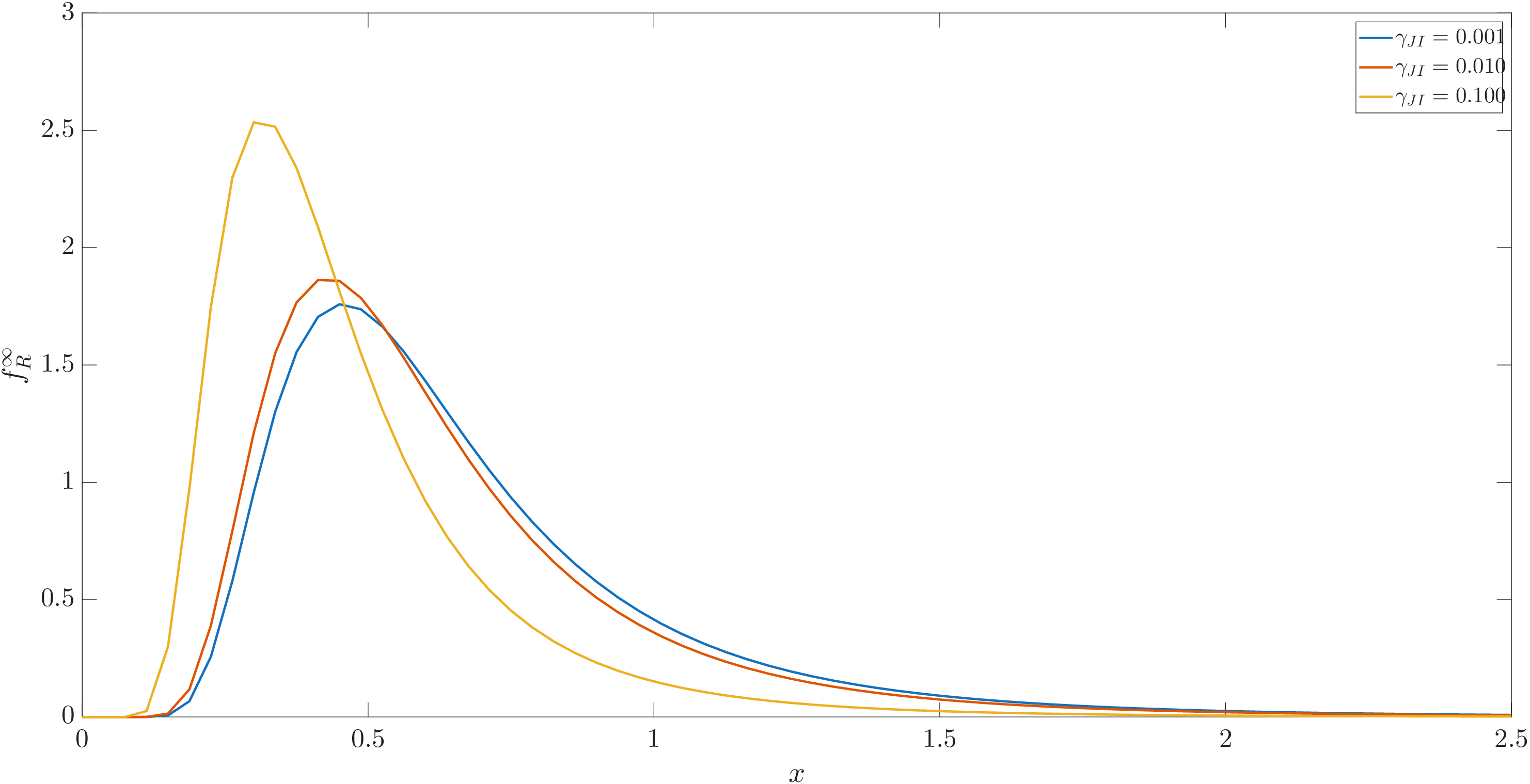}
    \includegraphics[width=70 mm]{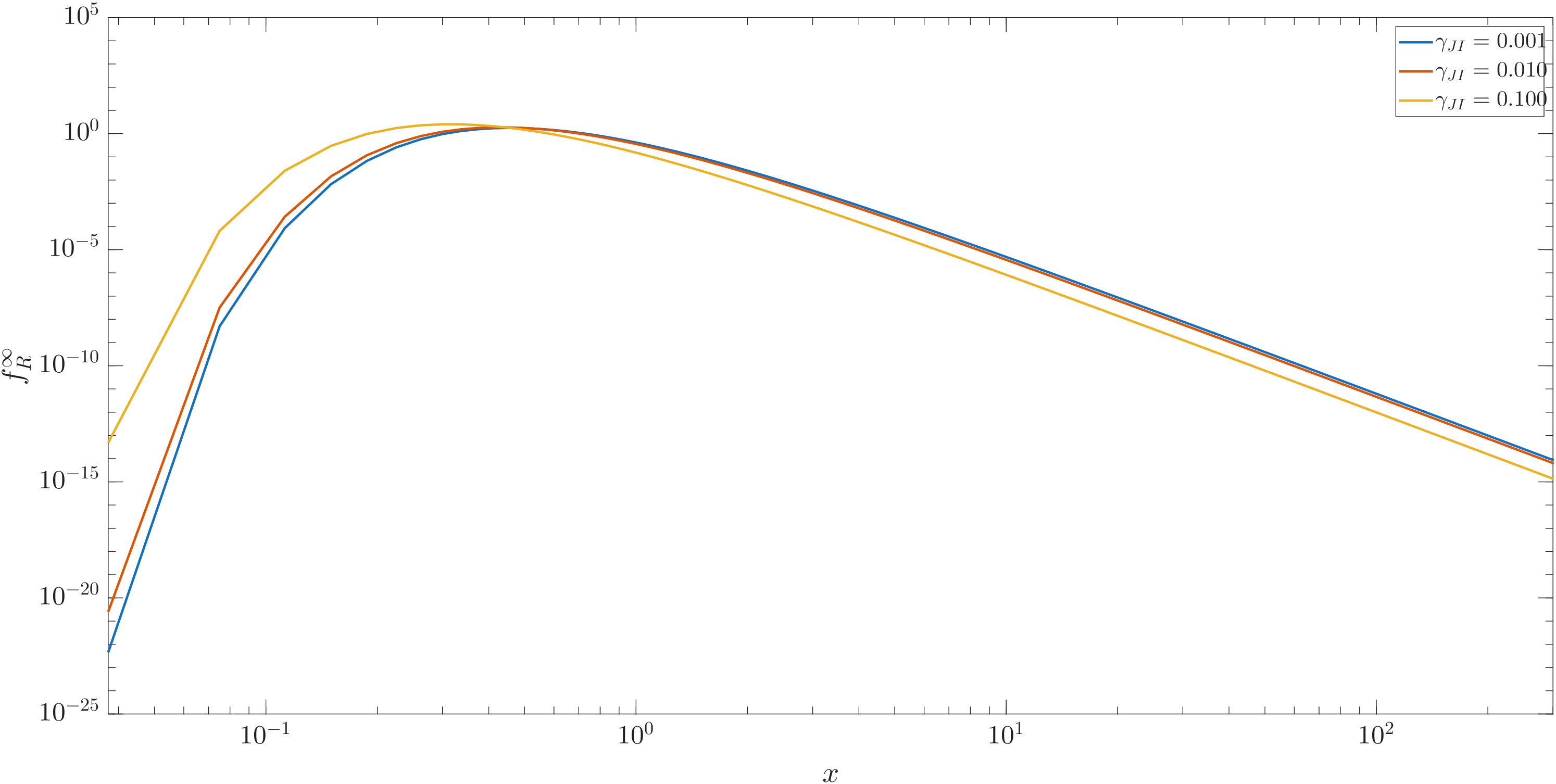}
    \caption{Different behaviors of the evolution for the resistant compartment, i.e. $R$, with respect to different values of the parameter $\gamma_{JI}$. From top to bottom and from left to right: the closed form of the mean $m_R$, the closed form of the variance $v_R$, and the equilibrium $f^{\infty}_R$, displayed using both linear and log–log scales.}
    \label{gammaJIpic}
\end{figure}

The effect of parameter $\gamma_{IR}$, i.e., the one of bad-treatment, and of the parameter $\gamma_{JI}$, i.e., the one of well-treatment, is shown in Figures \ref{gammaIRpic} and \ref{gammaJIpic}, respectively. We refer to the parameter values reported in Table~\ref{tab:parameters}, while varying $\gamma_{IR}\in [0.001,\, 0.01,\,0.10]$ and $\gamma_{JI}\in 
[0.001,\, 0.01,\, 0.10]$. It is worth noting that increasing the parameter $\gamma_{IR}$, associated with bad-treatment, corresponds to intensifying the resistant-strain pathogen, which is expected to impact the resistant compartment $R$. On the other hand, increasing $\gamma_{JI}$, associated with well-treatment, represents an improvement in effective therapies, enhancing the proper management of the resistant-strain pathogen. Roughly speaking, three scenarios can be identified, which we label as high-, medium-, and low-resistance regimes. These correspond, respectively, to larger values of $\gamma_{IR}$ (or, conversely, smaller values of $\gamma_{JI}$), intermediate configurations, and smaller values of $\gamma_{IR}$ (or, equivalently, larger values of $\gamma_{JI}$).

The numerical results clearly highlight the different roles played by $\gamma_{IR}$ and $\gamma_{JI}$ on the resistant compartment $R$.

Concerning Figure \ref{gammaIRpic}, increasing $\gamma_{IR}$ leads to a systematic growth of both the mean $m_R$ and the variance $v_R$. This behavior is consistent with the interpretation of $\gamma_{IR}$ as bad-treatment: ineffective therapies enhance the spread of the resistant-strain pathogen, resulting in a larger and more heterogeneous resistant population. From a modeling viewpoint, this corresponds to a stronger impact of resistance, reflecting a scenario in which the resistant compartment plays an increasingly dominant role in shaping the overall dynamics. This can be interpreted as the emergence and amplification of infections in particularly vulnerable settings, where susceptible individuals are exposed to higher risks due to environmental and clinical conditions. In this perspective, hospital environments represent a paradigmatic example, as they concentrate fragile patients, frequent contacts, and multiple transmission pathways, thereby enhancing the effect of resistant strains on the system. This situation is especially critical because it threatens to undermine the safety and effectiveness of routine medical procedures. As emphasized in reports by the World Health Organization, antimicrobial resistance can compromise standard clinical practices. This effect is further confirmed by the equilibrium distributions $f_R^\infty$. As $\gamma_{IR}$ increases, the peak of the distribution decreases while the tail becomes heavier, as clearly visible in the log-log scale. This indicates a higher dispersion and the presence of extreme values, coherently reflected in the increase of the variance. Overall, these features suggest a scenario in which antimicrobial resistance becomes harder to control due to the widespread presence of the resistant strain.

A complementary behavior is observed in Figure \ref{gammaJIpic} for the well-treatment parameter $\gamma_{JI}$. Increasing this parameter leads to a reduction of both the mean $m_R$ and the variance $v_R$, highlighting the beneficial role of well-treatment. Effective therapies limit the development and persistence of the resistant-strain pathogen, thus reducing both the size and the heterogeneity of the resistant compartment. The same trend is reflected in the equilibrium distributions: as $\gamma_{JI}$ increases, the profiles become more concentrated, with higher peaks and lighter tails. In the log-log representation, this corresponds to a faster decay, indicating a reduced probability of extreme values. This behavior confirms that well-treatment strategies effectively contain the spread of resistance, promoting a more controlled and less dispersed resistant population. However, the impact of $\gamma_{IR}$ on the resistant compartment appears to be more pronounced than that of $\gamma_{JI}$, at least within the considered parameter set. In particular, variations in $\gamma_{IR}$ lead to more significant changes in the shape and spread of the resistant profiles, suggesting that inadequate treatments play a dominant role in driving resistance. This asymmetry indicates that poorly administered treatments may have stronger consequences than effective ones: they not only contribute directly to the growth of the resistant population, but can also trigger new infection pathways, potentially involving previously unaffected susceptible individuals.

\section*{Conclusions}\label{sec:conclusions}
In this paper, we propose a novel model for microbial infection in the presence of antimicrobial resistance, aiming to characterise the long-time behaviour of the infection and quantify the impact of inappropriate antimicrobial use.
To this end, we introduce a kinetic framework describing the temporal evolution of the statistical distributions of population densities in the susceptible, infectious, recovered, and resistant compartments, driven by susceptible–infectious interactions. The model consists of a system of Boltzmann-type equations for binary interactions between susceptible and infectious individuals, supplemented by linear redistribution operators accounting for recovery and possible reinfection dynamics.
The model describes the evolution of the statistical densities of the population compartments and is linked to classical compartmental epidemiological models through the evolution of their mean values. These mean values satisfy a classical SIRS-type model, whose parameters are explicitly derived from the microscopic interaction rules.
Following the strategy recently employed in \cite{MTZ}, a grazing‑collision procedure shows that the Boltzmann system is well approximated by a coupled system of Fokker–Planck equations. This approximation leads to a significant simplification in the study of the large‑time behaviour of the population densities.
Within this framework, and following \cite{MTZ}, we show that the solution converges, in the so-called Energy distance \cite{auricchio2026kinetic,toscani_zanella26}, to the unique stable equilibrium of the mean‑field system, enhancing the quantification of the effects of inappropriate antimicrobial use on the distribution of resistant individuals. These results establish a multiscale bridge between compartmental models for antimicrobial resistance and individual-based modelling, suggesting new research directions in both the derivation of meanfield compartmental models and the modelling and control of many-agent systems.

Finally, by inspecting the emerging equilibrium distributions, we observe that the impact of transitions due to inappropriate treatments, $\gamma_{IR}>0$, is stronger than that characterising transitions from the infectious state of the resistant strain to the infectious state of the non-resistant strain, $\gamma_{JI}>0$. Within the framework of the proposed model, the misuse of antimicrobial treatments (i.e., bad treatment) appears to have a stronger effect on the emergence and spread of resistance than the corresponding beneficial impact of well-targeted treatment strategies.

\section*{Acknowledgements}
This work has been written within the activities of the GNFM
 group of INdAM (National Institute of High Mathematics). M.Z. acknowledges partial support
from the PRIN2022PNRR project No.P2022Z7ZAJ, European Union-NextGenerationEU. M.Z. acknowledges partial support by ICSC - Centro Nazionale di Ricerca in High Performance Computing, Big Data and Quantum Computing, funded by European Union-NextGenerationEU. 

\appendix

\section{Derivation of the Boltzmann-type system}
\label{sec:appendix}

In Section \ref{sec:kinetic} the elementary interactions leading to variations in the sizes of the various compartments of the population size 
were identified and discussed. Once the definition of these elementary processes  has been fixed,  the evolution of the statistical distribution functions $f_K(x,t), K\in \mathcal C$ can fruitfully follows by resorting to classical kinetic theory.

This is motivated by recent approaches where the description of social phenomena in a multi-agent system was successfully obtained by resorting to statistical physics, and, in particular, to methods borrowed from kinetic theory of rarefied gases \cite{ParTos-2013}. In this approach, the main goal of the mathematical modeling is to construct master equations of Boltzmann type, usually referred to as kinetic equations, describing the time-evolution of some characteristic of the agents, like wealth, opinion, knowledge, or, as in the case treated in this paper, of the description of the size distribution of the antimicrobial resistance in a population of agent's which are characterized by their belonging to one of the five  compartments $K \in \mathcal C$.

The Boltzmann description is heavily dependent from the type of interaction, which, as shown in Section \ref{sec:kinetic}, can be either linear or bilinear. In both cases, to avoid problems linked to the Jacobian of the transformations,  the weak formulation, which consists of studying the time evolution of the observable quantities, say $\varphi(x)$, is easier  to use. 

Hence, any linear interaction of type \fer{eq:redi1} moving individuals from the compartment $K$ to $L$, where $K,L \in \mathcal C$ leads to the linear kinetic equations
\begin{equations}\label{eqweak-lin}
    \frac{d}{dt}\int_{\mathbb{R}_+}\varphi(x)f_K(x,t)\,dx&=
    \int_{\mathbb{R}^2_+}\left[\varphi(x^*)-\varphi(x)\right]f_K(x,t)h_L(y,t)\,dxdy,  \\
    \frac{d}{dt}\int_{\mathbb{R}_+}\varphi(x)f_L(x,t)\,dx&=
    \int_{\mathbb{R}^2_+}\left[\varphi(x^*)-\varphi(x)\right]f_L(x,t)f_K(y,t)\,dx_Idy,
\end{equations}
which quantify in both compartments the negative and, respectively, positive variations of observable quantity $\varphi$ due to the interaction of individuals in the compartment $K$ with the background distribution $h_L$.

Also, the linear kinetic equations associated to the loss of immunity in the compartments $L=T,R$, as given by the elementary changes \fer{eqmicro71}, read
\begin{equations}\label{eq:loss}
    \frac{d}{dt}\int_{\mathbb{R}_+}\varphi(x)f_L(x,t)\,dx&=
    \int_{\mathbb{R}^2_+}\left\langle\varphi(x^*)-\varphi(x)\right\rangle f_L(x,t)\,dx,  \\
    \frac{d}{dt}\int_{\mathbb{R}_+}\varphi(x)f_S(y,t)\,dy&=
    \int_{\mathbb{R}^2_+}\left\langle\varphi(y^*)-\varphi(y)\right\rangle f_L(x,t)f_S(y,t)\,dx_Idy.
\end{equations}

Likewise, bilinear interactions of type \fer{eq:SK}--\fer{eq:K+} moving agents from the compartment $S$ to $L=I,J$ modify any observable quantity $\varphi(x)$ according to
\begin{equations}\label{eq:bolS1}
    \frac{d}{dt}\int_{\mathbb{R}_+}\varphi(x)f_S(x,t)\,dx&=\int_{\mathbb{R}^2_+}k(y)\langle\varphi(x^*)-\varphi(x)\rangle f_S(x,t)f_L(y,t)\,dxdy, \\
     \frac{d}{dt}\int_{\mathbb{R}_+}\varphi(y)f_L(y,t)\,dy&=\int_{\mathbb{R}^2_+}k(x)\langle\varphi(y^*)-\varphi(y)\rangle f_S(x,t)f_L(y,t)\,dxdy.
\end{equations}
where the kernel $k(\cdot)$ takes into account the frequency of the interactions by relating proportionally this frequency to the size of the interacting compartment. A classical choice, closely related to the structure of  interactions of type \fer{eq:SK}--\fer{eq:K+}, is to set
\be\label{eq:freq}
k(x) = 1+x.
\ee
Differently from \fer{eq:bolS1} the variation of individuals from $S$ to $J$, due to interactions with $R$, leads to the Boltzmann-like equations
\begin{equations}\label{eq:bolSR}
    \frac{d}{dt}\int_{\mathbb{R}_+}\varphi(x)f_S(x,t)\,dx&=\int_{\mathbb{R}^2_+}k(y)\langle\varphi(x^*)-\varphi(x)\rangle f_S(x,t)f_R(y,t)\,dxdy, \\
     \frac{d}{dt}\int_{\mathbb{R}_+}\varphi(z)f_J(z,t)\,dz&=\int_{\mathbb{R}^2_+}k(x)\langle\varphi(z^*)-\varphi(z)\rangle f_S(x,t)f_R(y,t)f_J(z,t)\,dxdydz.
\end{equations}
Consequently, grouping all interactions related to a given compartment $K$ with $K \in \mathcal C$, as defined in Sections \ref{sec:inte} and \ref{sec:linear}, we obtain that the densities $f_K(x,t), K\in\mathcal C$, satisfy, for any given smooth function $\varphi$, the five Boltzmann-type equations


\begin{equation}\label{eqweak1}
\begin{split}
    \frac{d}{dt}\int_{\mathbb{R}_+}\varphi(x_S)f_S(x_S,t)\,dx_S&=\int_{\mathbb{R}^2_+}k(x_I)\langle\varphi(x'_S)-\varphi(x_S)\rangle f_S(x_S,t)f_I(x_I,t)\,dx_Sdx_I\\
    &+\int_{\mathbb{R}^2_+}k(x_J)\langle\varphi(x''_S)-\varphi(x_S)\rangle f_S(x_S,t)f_J(x_J,t)\,dx_Sdx_J\\
    &+\int_{\mathbb{R}^2_+}k(x_R)\langle\varphi(x'''_S)-\varphi(x_S)\rangle f_S(x_S,t)f_R(x_R,t)\,dx_Sdx_R\\
    &+\int_{\mathbb{R}^2_+}\left(\varphi(x''''_S)-\varphi(x_S)\right) f_S(x_S,t)f_T(x_T,t)\,dx_Sdx_T\\
    &+\int_{\mathbb{R}^2_+}\left(\varphi(x'''''_S)-\varphi(x_S)\right) f_S(x_S,t)f_R(x_R,t)\,dx_Sdx_R
\end{split}    
\end{equation}

\begin{equation}\label{eqweak2}
\begin{split}
    \frac{d}{dt}\int_{\mathbb{R}_+}\varphi(x_I)f_I(x_I,t)\,dx_I&=\int_{\mathbb{R}^2_+}k(x_S)\langle\varphi(x'_I)-\varphi(x_I)\rangle f_S(x_S,t)f_I(x_I,t)\,dx_Sdx_I\\&
    +\int_{\mathbb{R}^2_+}\left(\varphi(x''_I)-\varphi(x_I)\right) f_I(x_I,t)f_J(x_J,t)\,dx_Jdx_I\\&
    +\int_{\mathbb{R}^2_+}\left(\varphi(x'''_I)-\varphi(x_I)\right) f_I(x_I,t)h_{IT}(y,t)\,dx_Idy \\&
    +\int_{\mathbb{R}^2_+}\left(\varphi(x''''_I)-\varphi(x_I)\right) f_I(x_I,t)h_{IR}(y,t)\,dx_Idy
\end{split}      
\end{equation}

\begin{equation}\label{eqweak3}
\begin{split}
    \frac{d}{dt}\int_{\mathbb{R}_+}\varphi(x_J)f_J(x_J,t)\,dx_J&=\int_{\mathbb{R}^2_+}k(x_S)\langle\varphi(x'_J)-\varphi(x_J)\rangle f_S(x_S,t)f_J(x_J,t)\,dx_Sdx_J\\&
    +\int_{\mathbb{R}^3_+}k(x_S)\langle\varphi(x''_J)-\varphi(x_J)\rangle f_S(x_S,t)f_R(x_R,t)f_J(x_J,t)\,dx_Sdx_Rdx_J\\&
    +\int_{\mathbb{R}^2_+}\left(\varphi(x'''_J)-\varphi(x_J)\right) f_J(x_J,t)h_{JI}(y,t)\,dx_Jdy\\&
    +\int_{\mathbb{R}^2_+}\left(\varphi(x''''_J)-\varphi(x_J)\right) f_J(x_J,t)h_{JR}(y,t)\,dx_Jdy
\end{split}      
\end{equation}

\begin{equation}\label{eqweak4}
\begin{split}
    \frac{d}{dt}\int_{\mathbb{R}_+}\varphi(x_T)f_T(x_T,t)\,dx_T&=\int_{\mathbb{R}^2_+}\left(\varphi(x'_T)-\varphi(x_T)\right) f_T(x_T,t)f_I(x_I,t)\,dx_Tdx_I\\&
    +\int_{\mathbb{R}_+}\langle\varphi(x''_T)-\varphi(x_T)\rangle f_T(x_T,t)\,dx_T
\end{split}    
\end{equation}

\begin{equation}\label{eqweak5}
\begin{split}
    \frac{d}{dt}\int_{\mathbb{R}_+}\varphi(x_R)f_R(x_R,t)\,dx_R&=\int_{\mathbb{R}^2_+}\left(\varphi(x'_R)-\varphi(x_R)\right) f_R(x_R,t)f_I(x_I,t)\,dx_Rdx_I\\&
    +\int_{\mathbb{R}^2_+}\left(\varphi(x''_R)-\varphi(x_R)\right) f_R(x_R,t)f_J(x_J,t)\,dx_Rdx_J\\&
    +\int_{\mathbb{R}_+}\langle\varphi(x'''_R)-\varphi(x_R)\rangle f_R(x_R,t)\,dx_R,
\end{split}    
\end{equation}

\subsection{The grazing interaction regime}\label{sec:grazing}

If interaction \fer{eq:SK} is modified according to \fer{eq:SKE}, where the parameter $\eps \ll 1$ we are in presence of a grazing interaction. The term \emph{grazing} translates the fact that the post-interaction value is very close to the pre-interaction value. 

In this situation, by expanding the smooth function $\varphi(x)$ in Taylor series up to the second order, we get
\begin{equations}
\langle \varphi(x_\eps^*) - \varphi(x) \rangle &= \varphi'(x) \langle x_\eps^* -x\rangle +\\
& +\frac 12 \varphi''(x) \varphi'(x) \langle (x_\eps^* -x)^2\rangle +
R_\eps(\varphi, x,x_\eps^*).
\end{equations}
Owing to \fer{eq:SKE}, it is immediate to show that, provided the third derivative of $\varphi$ is uniformly bounded, that
\[
\langle \varphi(x_\eps^*) - \varphi(x) \rangle  = \eps\left[ \varphi'(x) \beta\frac y{1+y} x + \frac 12 \varphi''(x) \sigma \frac y{1+y} x^2\right] +o(\eps).
\]
which, coupled with the time scaling $t \in t/\eps$, implies that only the coefficient of the term linear in $\eps$ survives. On the other hand, since the kernel $k(y) = 1+y$, any bilinear term of the form \fer{eq:bolS1}, as $\eps\to 0$, converges towards a Fokker--Planck type operator of type \fer{eq:FP-graz} in weak form.

The same procedure clearly applies to linear interactions. For completeness, we give below the complete set of scaling of the parameters of the Boltzmann system which lead to the system of five Fokker--Planck equations \fer{eq:FP-graz}
\begin{align*}
    &\beta_I \rightarrow \varepsilon \beta_I \qquad \beta_J \rightarrow \varepsilon \beta_J \qquad
    \sigma_{SI} \rightarrow \varepsilon \sigma_{SI} \qquad
    \sigma_{I} \rightarrow \varepsilon \sigma_{I}\\
    &\sigma_{SJ} \rightarrow \varepsilon \sigma_{SJ} \qquad
    \sigma_{J} \rightarrow \varepsilon \sigma_{J}\qquad
    \beta_R \rightarrow \varepsilon \beta_R\qquad
    \sigma_{SR} \rightarrow \varepsilon \sigma_{SR} \\
    &\sigma_{R} \rightarrow \varepsilon \sigma_{R}\qquad
    \gamma_{JI}\rightarrow\varepsilon\gamma_{JI}\qquad
    \gamma_{T}\rightarrow\varepsilon\gamma_{T}\qquad
    \gamma_{IR}\rightarrow\varepsilon\gamma_{IR}\\
    & \gamma_{R}\rightarrow\varepsilon\gamma_{R}\qquad
    \alpha_{T}\rightarrow\varepsilon\alpha_{T}\qquad
    \alpha_{R}\rightarrow\varepsilon\alpha_{R}\qquad
    \sigma_T \rightarrow\varepsilon\sigma_T\qquad
    \sigma_{RJ} \rightarrow\varepsilon\sigma_{RJ}.
\end{align*}

\section{Stability of equilibrium values}\label{appendix:stability}
We here provide the computations for the stability results concerning the equilibrium of the system of means \eqref{eqmean}. First, we write this system as follows: 

\begin{align}\label{eqmeannew}
\dot m_S &= -(\beta_I m_I + \beta_J m_J + \beta_R m_R)m_S + \alpha_T m_T + \alpha_R m_R,\\
\dot m_I &= \beta_I m_S m_I - (\gamma_T + \gamma_{IR})m_I + \gamma_{JI}m_J,\\
\dot m_J &= (\beta_J m_J + \beta_R m_R)m_S - (\gamma_{JI} + \gamma_R)m_J,\\
\dot m_T &= \gamma_T m_I - \alpha_T m_T,\\
\dot m_R &= \gamma_{IR}m_I + \gamma_R m_J - \alpha_R m_R.
\end{align}

The total mean is conserved:
\[
m_S + m_I + m_J + m_T + m_R = 1.
\]

The Jacobian of system \eqref{eqmeannew} is
\[
J =
\begin{pmatrix}
-\mu & -\beta_I m_S & -\beta_J m_S & \alpha_T & -\beta_R m_S + \alpha_R\\
\beta_I m_I & \beta_I m_S - (\gamma_T + \gamma_{IR}) & \gamma_{JI} & 0 & 0\\
\beta_J m_J + \beta_R m_R & 0 & \beta_J m_S - (\gamma_{JI} + \gamma_R) & 0 & \beta_R m_S\\
0 & \gamma_T & 0 & -\alpha_T & 0\\
0 & \gamma_{IR} & \gamma_R & 0 & -\alpha_R
\end{pmatrix},
\]
where
\[
\mu = \beta_I m_I + \beta_J m_J + \beta_R m_R,
\]
and it is evaluated at the endemic equilibrium
\[
(m_S^\infty,m_I^\infty,m_J^\infty,m_T^\infty,m_R^\infty).
\]
We introduce the notation
\[
P=\mu^\infty=\beta_I m_I^\infty+\beta_J m_J^\infty+\beta_R m_R^\infty,\quad 
Q=\frac{\gamma_{JI}m_J^\infty}{m_I^\infty}
+
\frac{\beta_R m_S^\infty m_R^\infty}{m_J^\infty},\quad
R=\beta_I m_S^\infty \gamma_{JI},
\]
and
\[
\alpha_1 = \alpha_T+\alpha_R,
\qquad
\alpha_2=\alpha_T\alpha_R.
\]
Using the endemic equilibrium identities for $\dot m_I$ and $\dot m_J$
\begin{equation}
\begin{split}
\beta_I m_S^\infty m_I^\infty + \gamma_{JI} m_J^\infty &= (\gamma_T+\gamma_{IR})m_I^\infty,\\
(\beta_J m_J^\infty + \beta_R m_R^\infty)m_S^\infty &= (\gamma_{JI}+\gamma_R)m_J^\infty,
\end{split}\end{equation}
the characteristic polynomial factors as
\[
\det(\lambda I - J) = \lambda p_4(\lambda),
\]
therefore $\lambda_1=0$ and $p_4$ is the degree-four polynomial
\[
p_4(\lambda)=\lambda^4+c_3\lambda^3+c_2\lambda^2+c_1\lambda+c_0,
\]
with coefficients
\begin{equation}
\begin{split}
c_3 &= \alpha_1 + P + Q,\\
c_2 &= \alpha_2 + \alpha_1(P+Q) + PQ + R,\\
c_1 &= \alpha_2(P+Q) + \alpha_1(PQ+R),\\
c_0 &= \alpha_2(PQ+R).
\end{split}
\end{equation}
The coefficients above satisfy the identity
\[
p_4(\lambda) =
(\lambda^2 + A\lambda + B)
(\lambda^2 + \alpha_1\lambda + \alpha_2),
\]
where
\[
A=P+Q,
\qquad
B=PQ+R.
\]
The second factor further factors as follows
\[
\lambda^2+\alpha_1\lambda+\alpha_2
=
(\lambda+\alpha_T)(\lambda+\alpha_R),
\]
yielding two exact eigenvalues
\[
\lambda_2=-\alpha_T,
\qquad
\lambda_3=-\alpha_R.
\]
The remaining two eigenvalues $\lambda_{4,5}$ satisfy 
\begin{equation}
\label{eq:quadratic}
\lambda^2+A\lambda+B=0.
\end{equation}
Since all parameters and all components of the endemic equilibrium are strictly positive
\[
P>0,
\qquad
Q>0,
\qquad
R>0,
\]
we obtain
\[
A=P+Q>0,
\qquad
B=PQ+R>0.
\]
Finally, solving \eqref{eq:quadratic} we get
\[
\lambda_{4,5}
=
\frac{-A\pm\sqrt{A^2-4B}}{2}.
\]
In particular, we get
\begin{itemize}
\item[$i)$] If $A^2\ge4B$: both roots are real and negative since $A>0$ and $B>0$.
\item[$ii)$] If $A^2<4B$: the roots are complex conjugates with real part $-A/2<0$.
\end{itemize}
In both cases, we have obtained that $\mathrm{Re}(\lambda_{4,5})<0$, and therefore the endemic equilibrium
\[
(m_S^\infty,m_I^\infty,m_J^\infty,m_T^\infty,m_R^\infty)
\]
is locally asymptotically stable. The spectrum of the Jacobian $J$ evaluated at this equilibrium is

\[
\lambda_1=0,
\quad
\lambda_2=-\alpha_T,
\quad
\lambda_3=-\alpha_R,
\]

\[
\lambda_{4,5}
=
\frac{-(P+Q)\pm\sqrt{(P+Q)^2-4(PQ+R)}}{2},
\]
where
\[
P=\beta_I m_I^\infty+\beta_J m_J^\infty+\beta_R m_R^\infty,\quad 
Q=
\frac{\gamma_{JI}m_J^\infty}{m_I^\infty}
+
\frac{\beta_R m_S^\infty m_R^\infty}{m_J^\infty},\quad 
R=\beta_I m_S^\infty \gamma_{JI}.
\]
All remaining eigenvalues have strictly negative real part, so the endemic equilibrium is locally asymptotically stable on the invariant hyperplane
\[
\{m_S+m_I+m_J+m_T+m_R=1\}.
\]

\bibliographystyle{plain} 
\bibliography{biblio}

\end{document}